\documentclass[10pt, conference]{IEEEtran}
\usepackage{url}
\usepackage{cite}
\usepackage{indentfirst}
\IEEEoverridecommandlockouts
\usepackage{graphicx}
\usepackage{svg}
\usepackage{amsmath,amssymb,amsfonts}
\usepackage{algorithmic}
\usepackage{mathtools}
\usepackage{algorithm}
\usepackage{graphicx}
\usepackage{float}
\usepackage{textcomp}
\usepackage{subfigure} 
\usepackage{mwe}
\usepackage{textcomp}
\usepackage{float}
\usepackage{wrapfig}
\usepackage{rotating}
\usepackage{tikz}
\usepackage{textcomp}
\usepackage{mwe}
\usepackage{multicol}
\usepackage{multirow}
\usepackage{graphicx}
\usepackage{float}
\usepackage{amssymb}
\usepackage{gensymb}
\usepackage{array}
\usepackage{enumitem}
\usepackage{fancyhdr}
\usepackage{dirtytalk}
\usepackage{flexisym}
\usepackage{anyfontsize}
\usepackage[T1]{fontenc}
\usepackage{booktabs}

\usepackage[]{caption}

\usepackage{amsmath,amssymb,amsfonts}
\usepackage{algorithmic}
\usepackage{graphicx} 
\usepackage{textcomp}
\usepackage{xcolor}
\usepackage{rotating}
\usepackage{tikz}
\usepackage[hidelinks]{hyperref}

\newcommand{\linebreakand}{%
      \end{@IEEEauthorhalign}
      \hfill\mbox{}\par
      \mbox{}\hfill\begin{@IEEEauthorhalign}
    }

\newcommand{\ourmethod}{\textit{mmDrive}\;}

\begin{document}
\title{\ourmethod: mmWave Sensing for Live Monitoring and On-Device Inference of Dangerous Driving
}

\author{\IEEEauthorblockN{Argha Sen\IEEEauthorrefmark{1}, Avijit Mandal\IEEEauthorrefmark{2}, Prasenjit Karmakar\IEEEauthorrefmark{3}},
\IEEEauthorblockN{Anirban Das\IEEEauthorrefmark{4}, Sandip Chakraborty\IEEEauthorrefmark{6}}
\IEEEauthorblockA{Department of Computer Science and Engineering,
Indian Institute of Technology Kharagpur\\
Email: \IEEEauthorrefmark{1}arghasen10@gmail.com,
\IEEEauthorrefmark{2}avijitmandal2001@gmail.com,
\IEEEauthorrefmark{3}prasenjitkarmakar52282@gmail.com\\
\IEEEauthorrefmark{4}anirband@iitkgp.ac.in,
\IEEEauthorrefmark{6}sandipchkraborty@gmail.com}
}

\maketitle

\begin{abstract}
Detecting dangerous driving has been of critical interest for the past few years. However, a practical yet minimally intrusive solution remains challenging as existing technologies heavily rely on visual features or physical proximity. With this motivation, we explore the feasibility of purely using mmWave radars to detect dangerous driving behaviors. We first study characteristics of dangerous driving and find some unique patterns of range-doppler caused by $9$ typical dangerous driving actions. We then develop a novel Fused-CNN model to detect dangerous driving instances from regular driving and classify $9$ different dangerous driving actions. Through extensive experiments with $5$ volunteer drivers in real driving environments, we observe that our system can distinguish dangerous driving actions with an average accuracy of $97(\pm 2)\%$. We also compare our models with existing state-of-the-art baselines to establish their significance. 
\end{abstract}


\section{Introduction}
World Health Organization (WHO) reports that an alarming $1.2$ million people die every year globally due to road accidents~\cite{world2020road}; dangerous driving remains one of the major causes (around $45\%$ of the cases) of these accidents. Notably, there have been significant technology-backed advancements toward monitoring dangerous driving in real-time. With commercial attempts such as \textit{Advance Driver Assistance Systems (ADAS)}~\cite{shaout2011advanced}, and numerous research works~\cite{yin2017automatic, jiang2021smart}, the domain is well-studied. For instance, recent times observe a significant drift towards computer vision~\cite{yang2020driver} and body-mounted wearable-based~\cite{cao2022towards} approaches for driving behavior monitoring. However, vision-based approaches are challenged by privacy concerns among individuals, particularly for the public and shared vehicles. Further, the detection accuracy depends on several factors~\cite{wang2021survey} such as lighting conditions, the orientation of the camera, etc. Similarly, wearable-based approaches are difficult to generalize as the signatures from a single age group do not necessarily align seamlessly with a different age group~\cite{kundinger2020feasibility}. Moreover, it is not easy to ensure a driver always equips themselves with the required wearable devices. At this point, we raise the following fundamental question. \textit{How can we develop a system for monitoring dangerous driving in a compact and pervasive manner, such that the community can adopt it at large? At the same time, can an alternative modality be leveraged effectively to address this problem?} We proceed with the notion of answering these. 



Recently, we observed an imminent paradigm shift towards 5G technology which builds on top of millimeter wave (mmWave)-based communication~\cite{hur2016proposal}. As more and more devices are introduced in this ecosystem, the mmWave hardware gets integrated into a large variety of devices, making this technology a pervasive platform. This observation metamorphose into a critical question: \textit{can we leverage the physics of mmWave to capture the driver's driving behaviors corresponding to dangerous driving behavior?} A mmWave radar~\cite{li2020capturing} can effectively measure parameters like distance, velocity, etc., of an object. This idea has been exploited to solve diverse range of problems, such as human activity recognition~\cite{wang2021m}, gesture recognition~\cite{palipana2021pantomime}, vital sign detection~\cite{yang2016monitoring} and even voice reconstruction~\cite{liu2021wavoice}, which involve positioning and movement estimations. Based on this information, in this work, we propose \textit{\ourmethod, that uses a Frequency-modulated continuous-wave (FMCW) mmWave radar to monitor dangerous driving behaviors from a driver's perspective}. As soon as a case of distracted driving is detected, \ourmethod can be used to take immediate action, such as warning the driver or notifying nearby vehicles with appropriate messages. 


\noindent\textbf{Advantages over Existing Approaches:} We also observe that leveraging mmWave for monitoring dangerous driving has several advantages over existing state-of-the-art approaches, including the following. (i) An mmWave radar can monitor a driver's movements directly rather than indirect observations such as vehicle states and its kinematics. (ii) Unlike cameras, an mmWave radar has a minimal invasion of privacy as it does not capture visual features of the environment. (iii) mmWave radar measures passively, and does not restrict an individual's regular movements and activities. Unlike wearables, mmWave sensing does not require a user to mount, or carry any device. (iv) With an mmWave radar, micro-movements such as \textit{yawning} can also be detected~\cite{sen2023mmassist,liu2022remote}. This could be crucial for determining a sleepy state of a driver. (v) mmWave can penetrate potentially light-occluding entities such as clothing~\cite{zhao2020heart}, and therefore can be used even with the driver wearing a face mask. However, beyond these advantages, \ourmethod must overcome a set of hurdles to be practically usable.


\noindent\textbf{Challenges:} It is not straightforward when it comes to using a mmWave radar to monitor the activities within a moving vehicle; several challenges must be addressed for a functionally effective system, as follows. \textit{Firstly}, the environment within a car is quite noisy. Many movements from the objects inside a car, as well as the road and traffic conditions, can directly impact the mmWave signals. \textit{Secondly} A car can have multiple passengers. However, only the movements specific to the driver are the key to monitoring dangerous driving scenarios; thus, only the drivers' movements need to be separated. \textit{Thirdly}, signatures captured by the FMCW radar from sudden jerks, such as in case of road bumps, potholes, etc. have significant variation patterns and therefore need to be eliminated as they do not bear relevant information.

\noindent\textbf{Our Contributions:} Owing to the above challenges, this paper develops a pervasive system for non-intrusive live monitoring of dangerous driver behaviors using mmWave sensing. In contrast to the existing works~\cite{jiang2021driversonar} that use acoustic sonar to detect only a few simple driving behaviors, we consider $9$ different complex macro and micro-level activities that can be potentially fatal during driving. The contributions of this paper are as follows.

\noindent \textbf{(1) Defining dangerous driving and identifying signatures to detect them.} We define \textit{three} actions typically caused due to fatigue/drowsiness of the driver (i.e., \textit{nodding}, \textit{yawning}, \textit{steering anomaly}), and \textit{six} actions indicating driver distraction (i.e., \textit{drinking/eating}, \textit{turning back}, \textit{picking up/drops}, \textit{fething forward}, \textit{speaking on mobile}, \textit{turning heads to talk to the passengers}). Altogether, these nine actions are potentially dangerous and should be avoided while driving. We leverage a single Commercial Off-The-Shelf (COTS) FMCW mmWave radar. Through thorough pilot experiments, we analyze a set of signal features, namely \textit{range doppler}, \textit{range profile}, and \textit{noise profile} for differentiating these activities from the regular driving actions. We highlight the requirements of combining spatial and temporal variations of these features to identify dangerous driving behaviors.\\
\noindent \textbf{(2) Development of an end-to-end pipeline for driving behavior classification over noisy data.} Through real-world experiments, we observe that mmWave data is affected by signal noises introduced from bad road conditions such as bumps, potholes, etc. We leverage IMU sensor data to detect and counter such road-induced noises. Further, we propose a novel Fused-CNN classifier to detect dangerous versus normal driving. Notably, \ourmethod not only differentiates dangerous driving from normal driving, but it also classifies the nine different instances of dangerous driving behavior, as mentioned above. In order to reduce the computational overhead and energy consumption, \ourmethod classifies among the nine potentially harmful driving actions only upon detecting an instance of dangerous driving.\\
\noindent \textbf{(3) Implementation and evaluation over on-the-field data.} We implemented our proposed Fused-CNN-based driver behavior model and compared it with a random forest and VGG-16-based baseline. We further compare our system with an acoustic modality-based approach~\cite{jiang2021driversonar} to demonstrate the superior accuracy of mmWave-based sensing. To reproduce our results, we open-source our implementation and a sample subset of our dataset: \url{https://github.com/arghasen10/mmdrive.git}. Through a thorough deployment of the device and field experimentation, we collected $20$ hours of driving data (mmWave, IMU, and dashcam video for ground truth) from $5$ users using three vehicles -- two sedans and one SUV. 

To the best of our knowledge, \ourmethod is the first of its kind that can capture nine dangerous driving actions using a mmWave radar. Moreover, it offers significantly higher accuracy than the closest baseline using FMCW Sonar~\cite{jiang2021driversonar} and is preferable over potentially privacy-invasive video-based techniques simultaneously. Our experimental results show that \ourmethod achieves an average accuracy of $90(\pm 0.5)\%$ in classifying dangerous driving from regular driving and $97(\pm 2)\%$ in detecting individual dangerous driving actions.

\section{Related Work}
Existing literature on monitoring dangerous driving behaviors branches out in multiple directions. Here, we summarize the most notable works in each of these directions.
\noindent\textit{Orientation and abnormality-based approaches:} 
These older approaches~\cite{ersal2010model,dai2010mobile} assess the orientation of the vehicle defined by a set of parameters such as position, velocity, acceleration, etc., and subsequently exploit them to \textit{indirectly} monitor dangerous driving. The assumption here is that the abnormality in the driving behavior causes irregular patterns in these parameters, eventually inferring dangerous driving. However, several uncontrollable elements, such as weather, traffic, etc., could also cause irregular driving patterns and not necessarily signify dangerous driving. 

\noindent\textit{Vision-based approaches:} Vision-based approaches~\cite{cyganek2014hybrid, borghi2017poseidon} leverage computer vision on RGB camera~\cite{dua2020dgaze}, thermal imaging~\cite{kajiwara2021driver}, and Infrared (IR) camera~\cite{yan2011robust} to detect abnormal driving activities by directly focusing on the driver. The captured images within the environment are processed to determine movements, including facial features such as eye movements, talking, and yawning, as well as movements of other body parts~\cite{hoang2016multiple} such as head movements, hand movements, etc. These activities, in turn, signify patterns of inattentive or dangerous driving behaviors. However, such vision-based approaches invade privacy~\cite{zhang2016privacy}, limiting their adoption as a practical solution primarily for public vehicles.


\noindent\textit{Wearable-based approaches:} Considering the pervasiveness and effectiveness of wearable devices, researchers have leveraged them to sense body kinematics and other body-specific signatures. Patterns in IMU (Inertial Measurement Unit) information, EEG (ElectroEncephaloGram)~\cite{cisotto2018joint}  signals, heart rate, etc., bear various activity-specific signatures, and thus are used to assess abnormal and dangerous driving behaviors eventually. However, such signatures often vary between different demographics (e.g., age groups), making it difficult to make such a system generic. Additionally, wearables must be explicitly carried by the users and may have invasive footprints (e.g., EEG sensor fitted on the body) interfering with individuals’ day-to-day movements.

\noindent\textit{Acoustics-based approaches:} Researchers have also explored acoustics for studying distracted and dangerous driving scenarios, which leverages doppler shifts~\cite{xu2017er, xie2019d} as well as FMCW chirps~\cite{jiang2021driversonar}. However, environmental noise has a significant influence~\cite{bai2020acoustic} over such acoustic-based sensing. Even with pervasive devices such as smartphones being used for acoustic sensing, it is worth noting that different devices have different sensitivity patterns~\cite{hromadova2022frequency} for audio signals that impact the effectiveness of such systems. Other factors, such as the impact of location and orientation of the sensor, and even privacy issues, are the cause of concern for acoustic sensing.


\section{Background and Observations}\label{sec:back}

\begin{figure}[t]
    \centering
    \subfigure[]{
    \includegraphics[width=0.3\columnwidth]{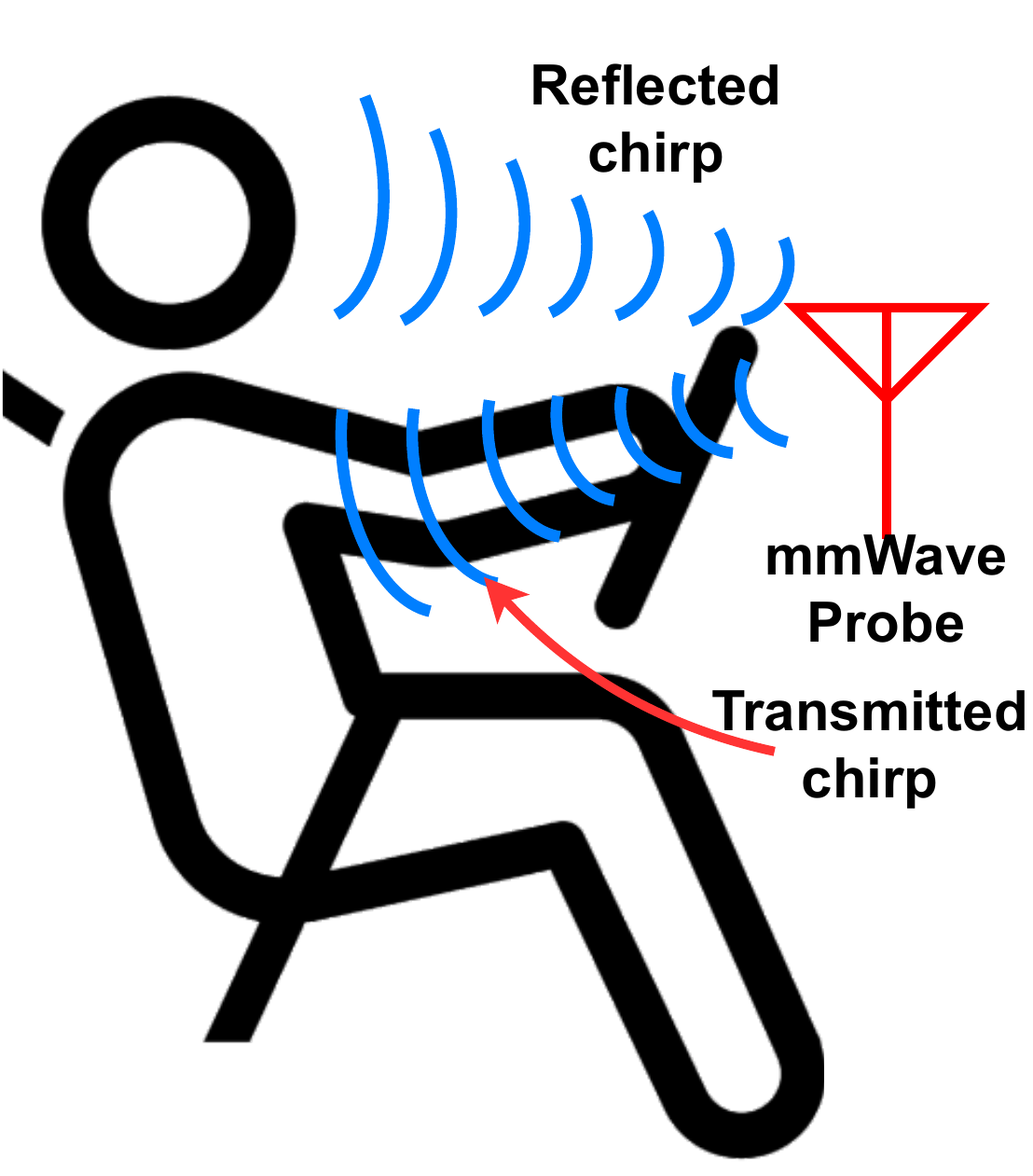}
    \label{fig:radar_placement}}
    \subfigure[]{
    \includegraphics[width=0.54\columnwidth]{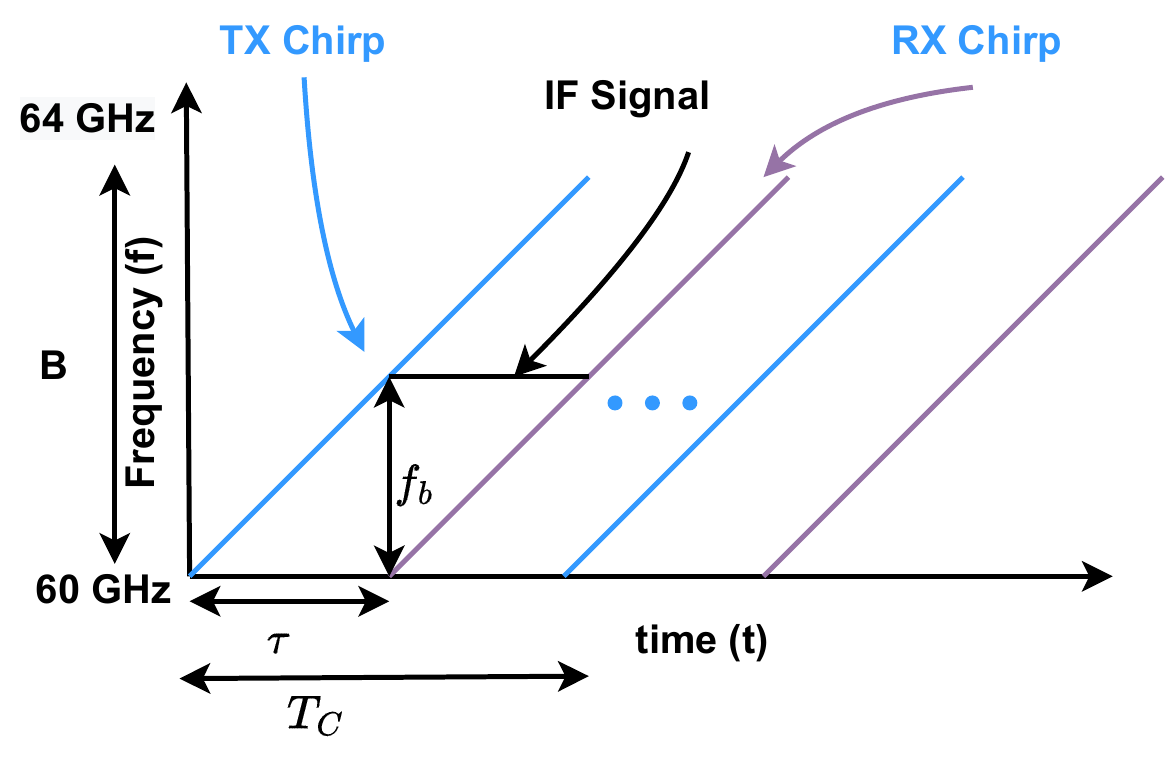}
    \label{fig:fmcw_working}}
    \caption{Working principle of an FMCW Radar - (a) Radar Placement (b) Frequency of multiple chirps vs. time} 
        \label{fig:mmwavefunction}
  \end{figure}

This section first introduces the technical aspects of an FMCW mmWave radar that are relevant to this work. After that, we present our observations concerning dangerous driving behaviors that can be mapped to these factors.
\subsection{Preliminaries}
\label{sec:prelims}
A generic FMCW radar uses a linear \textit{`chirp'} (continuous waves in the form of periodic signals) or swept frequency transmission, which is reflected by the obstacles in the environment. The radar then performs the \textit{dechirping} by mixing the transmitted signal with the reflected signal. The resulting \textit{Intermediate frequency} signal (IF) then undergoes a set of processing to extract the needed information.

\noindent\textit{Estimating the position of an object}
With a transmitted \textit{chirp} ($T_X$) and the corresponding received \textit{chirp} ($R_X$) and with a transmission time $T_C$, the distance $d$ of the body from the radar, which is causing the reflection, can be calculated as,
\begin{equation}\label{eq:range}
    d = \frac{c}{2} . \frac{T_C}{B} . f_b
\end{equation}

Here, $f_b$, the beat frequency, is the frequency difference between $T_X$ and $R_X$ chirps, $c$ is the speed of light, and the fractional component, $\frac{T_C}{B}$ represents the slope, $S$ of the FMCW chirp.
A \textit{Range Fast Fourier Transform} (Range-FFT) is needed first on the IF signal to extract all the reflectors (marked by frequency peaks). Eq.~\ref{eq:range} is then used to measure the distance of a reflector. Finally, the radar captures the received power at different \textit{range bins} and generates a one-dimensional array signifying range profile. \\
\noindent\textit{Estimating the movements of an object:}
An FMCW radar transmits $N$ number of chirps in a continuous sequence, each separated by a transmission time of $T_C$. When a body movement exists, the range-FFT corresponding to the chirps will have peaks in the exact location but with a \textit{phase change}. Let $v$ be the velocity at which the body moved, then the measured phase difference between two successive $R_X$ chirps corresponding to a motion of $v\times T_C$ can be stated as,
\begin{equation}
    \Delta \phi = \frac{4\pi (v \times T_C)}{\lambda}
\end{equation}
At this point, a second FFT, called \textit{Doppler-FFT}, can be performed on these $N$ phases to assess the body's movement. The radar captures the range and velocity information in a 2D matrix called a \textit{range-doppler} heatmap. 

\noindent\textit{Assessing the noise profile:} Doppler information at different range bins helps calculate the \textit{noise profile}. The noise profile is in the same format as \textit{range profile}, but the profile is at the maximum Doppler bin (maximum body movements). 

\ourmethod leverages \textit{range profile}, \textit{range-Doppler} heatmap and the \textit{noise profile} to capture signatures of \textit{driver activities}. For this, we perform a pilot study across different driver's behavior leveraging these factors.


\subsection{Pilot Study}\label{sec:pilot_study}
To design \ourmethod, the first question we come across is, \textit{how to define the problem of irregular and dangerous driving behavior}? Following the definitions of dangerous driving actions from the literature~\cite{3478084}, we extend it to accommodate a broader range of fine-grained activities that are indicators of dangerous driving. We primarily group these activities into two basic categories: (1) actions caused by fatigue/drowsiness and (2) actions caused by distraction. 

\begin{figure}[t]%
	\centering
	\subfigure[]{%
		\label{fig:nod}%
		\includegraphics[width=0.19\columnwidth]{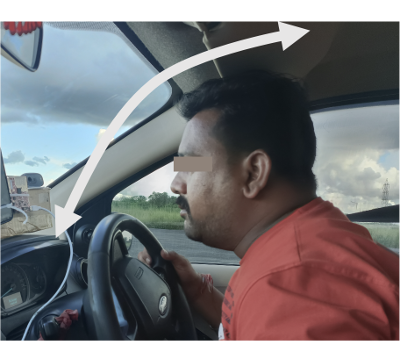}}\hfil%
	\subfigure[]{%
		\label{fig:yawn}%
		\includegraphics[width=0.19\columnwidth]{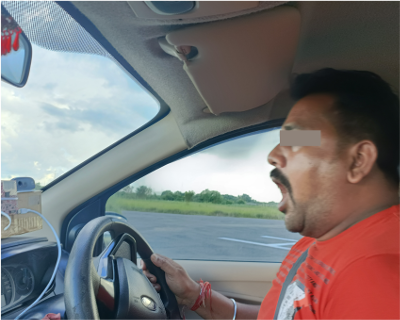}}\hfil%
	\subfigure[]{%
		\label{fig:steer}%
		\includegraphics[width=0.19\columnwidth]{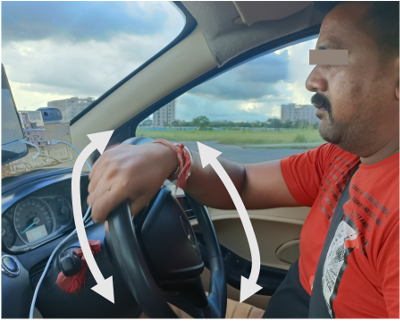}}\hfil%
	\subfigure[]{%
		\label{fig:drink}%
		\includegraphics[width=0.19\columnwidth]{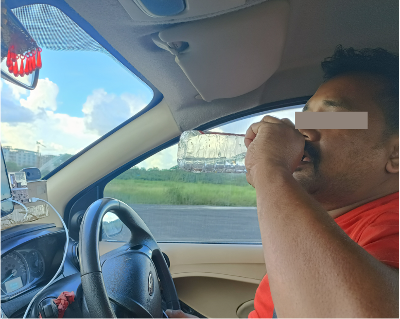}}\hfil%
	\subfigure[]{%
		\label{fig:turnbck}%
		\includegraphics[width=0.19\columnwidth]{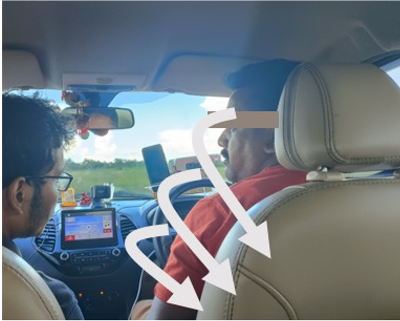}}\hfil%
	\subfigure[]{%
		\label{fig:pick}%
		\includegraphics[width=0.19\columnwidth]{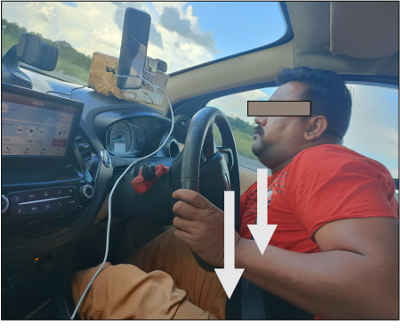}}\hfil%
	\subfigure[]{%
		\label{fig:fetch}%
		\includegraphics[width=0.19\columnwidth]{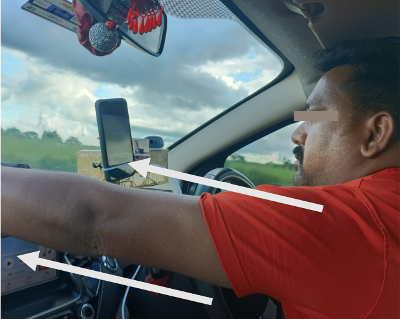}}\hfil%
	\subfigure[]{%
		\label{fig:mobile}%
		\includegraphics[width=0.19\columnwidth]{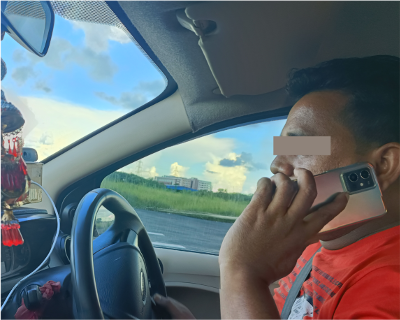}}\hfil%
	\subfigure[]{%
		\label{fig:talk}%
		\includegraphics[width=0.19\columnwidth]{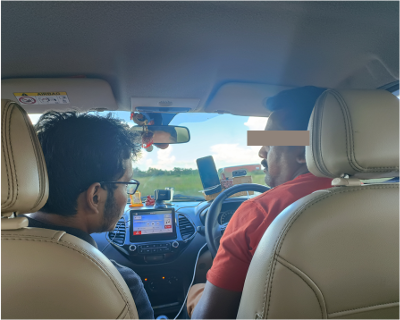}}%
	\caption{Dangerous driving activities -- (a) Nodding, (b) Yawning, (c) Steering anomaly, (d) Drinking, (e) Talking to the rear passenger, (f) Picking a drop, (g) Fetching from the dash, (h) Using mobile, (i) Talking sideways}
	\label{fig:driving_activities}
\end{figure}

\noindent\textit{Actions caused by drowsiness:} This group of actions typically occurs when the driver is fatigued or drowsy and could potentially cause an accident.
\begin{itemize}[leftmargin=*]
    \item \textit{Nodding:} When a driver is feeling drowsy or going to fall asleep, he/she typically makes a fast periodic bowing motion of the head in front and back direction (Fig.~\ref{fig:nod}).
    \item \textit{Yawning:} A standard action when the driver is fatigued is yawning. The two acts that make up a yawn include widely opening the mouth for inspiration, lifting the head due to the reactive force of expiration, and finally lowering the head to its normal position. (Fig.~\ref{fig:yawn})
    \item \textit{Anomaly in steering:} A typical driver often turns the steering wheel smoothly and uniformly, even on turns. However, a drowsy driver could stop adjusting the steering wheel for longer due to sleepiness. As the vehicle moves laterally, the driver may need to move the wheel sharply to adjust, which distinctively has movement patterns not observed during the normal driving scenario (Fig.~\ref{fig:steer}).
\end{itemize}

\noindent\textit{Actions caused by distractions:} Distracted driving typically involves activities that signify a lack of focus on driving and may result in dangerous situations. Here, we enlist a few such activities which we aim to capture using \ourmethod.
\begin{itemize}[leftmargin=*]
    \item \textit{Drinking or Eating:} An event of eating or drinking while driving might cause a minor forward tilt followed by picking up some items. In a typical scenario, one hand is momentarily taken off the wheel to grab the item and put it in the mouth (Fig.~\ref{fig:drink}). 
    \item \textit{Turning back:} A driver may turn around to talk to a rear passenger, check on some items in the backseat, or care for their children sitting behind. This, for obvious reasons, takes away their focus from driving. The typical movements observed were turning the head and twisting the upper part of the body in some cases (Fig.~\ref{fig:turnbck}). 
    \item \textit{Picking up drops:} Another situation is when a driver gets distracted when they intend to pick up a dropped item from the car floor while driving. This usually involves the movement of a hand and the upper part of the body downwards (Fig.~\ref{fig:pick}). 
    \item \textit{Fetching forward} A driver may intend to fetch something from the car dashboard while driving (Fig.~\ref{fig:fetch}). The movements involved typically include leaning forward along with necessary hand movements.
    \item \textit{Using Mobile Phone:} Using a mobile phone is a definite cause of distraction. A driver may pick up a call, text, or use an application by lifting the mobile phone (Fig.~\ref{fig:mobile}). 
    \item \textit{Talking to the front-passenger:} 
    The driver may intend to speak with the adjacent passenger (Fig.~\ref{fig:talk}). The usual movements involve turning the head and some mouth movements as the driver speak.
\end{itemize}

\noindent\textit{Other miscellaneous movements:} Apart from the movements highlighted above, the environment within a vehicle can also involve movements from elements such as \textit{road-bumps}. Even though these movements do not bear any information relevant to dangerous driving, the patterns are significant enough to confuse the classification model. Therefore, these movements need to be identified and eliminated accordingly.


Notably, the highlighted movements involved in the above cases are not transitory but last for a short while. We next discuss the observed signatures for the events mentioned above. In each case, to capture the signatures, an \texttt{AWR6843ISK}~\footnote{\url{https://www.ti.com/tool/AWR6843ISK} (Accessed: \today)} mmWave FMCW radar is placed on the car dashboard in front of the driver, as illustrated in Fig.~\ref{fig:radar_placement}. 

\begin{figure}[!ht]
    \centering
    \includegraphics[width=0.4\textwidth]{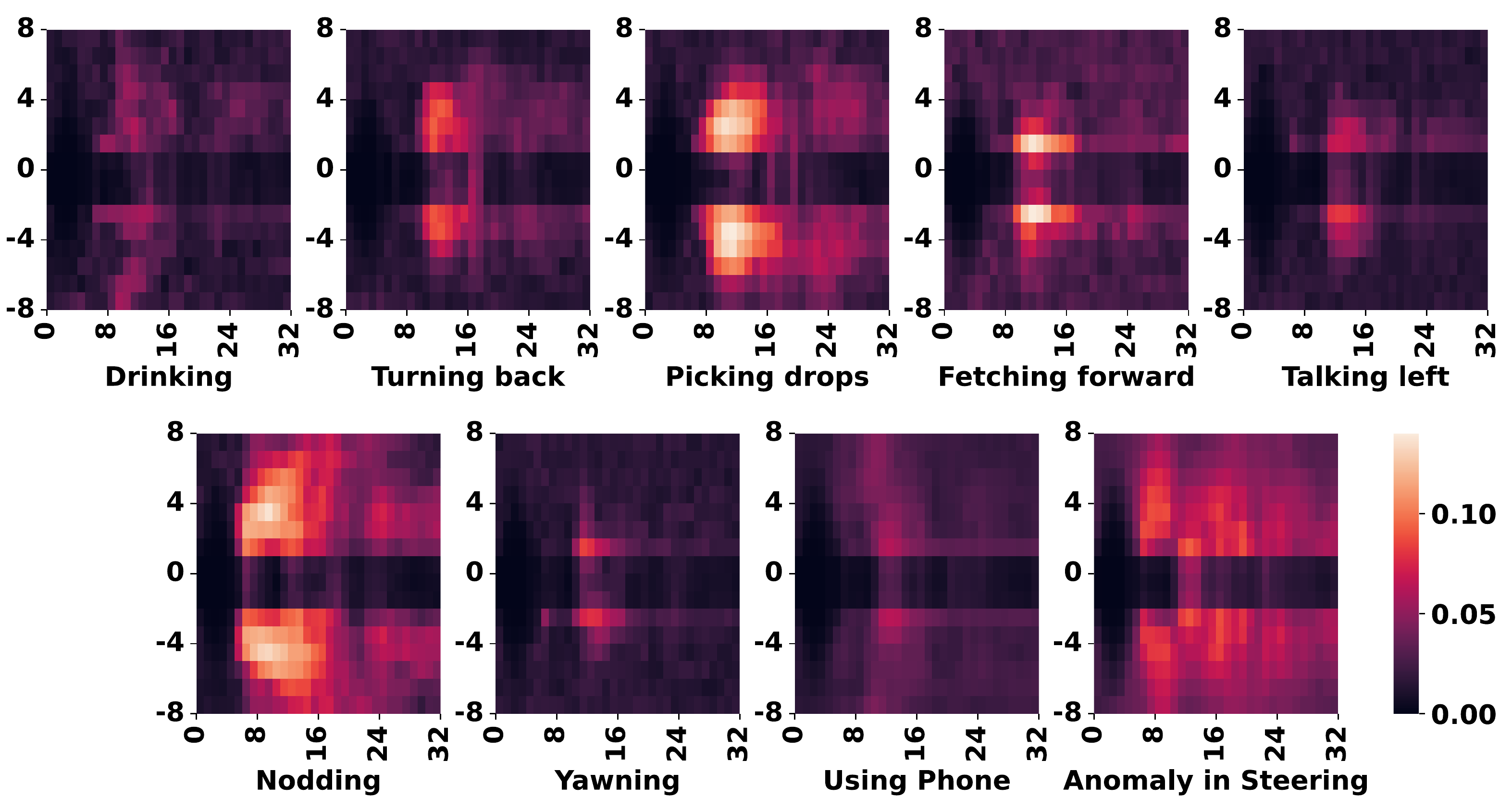}
    \caption{Range-doppler heatmaps for driving actions}
    \label{fig:raw_range_doppler}
\end{figure}

\begin{figure*}[]
    \centering
    \subfigure[]{
    \includegraphics[width=0.31\textwidth]{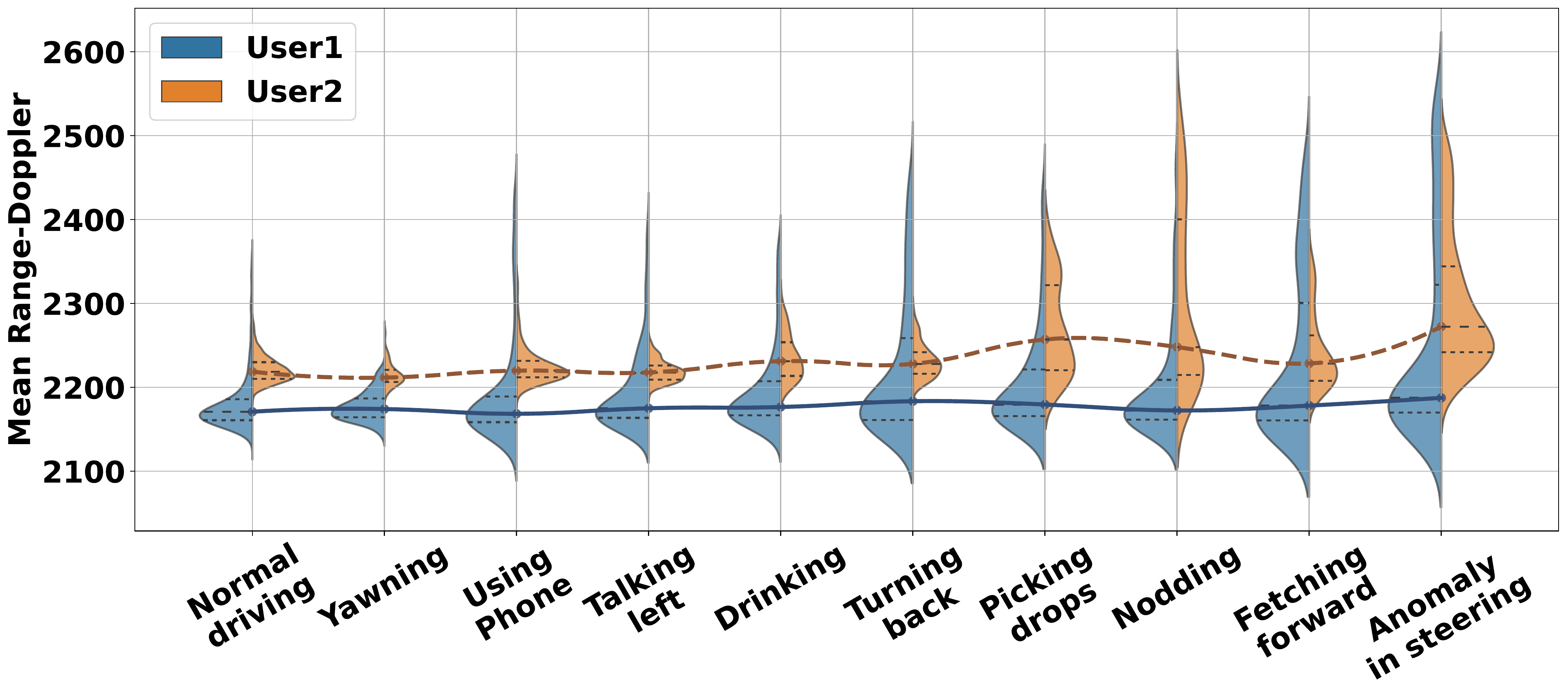}
            \label{fig:doppler_violin}}
    \subfigure[]{
    \includegraphics[width=0.31\textwidth]{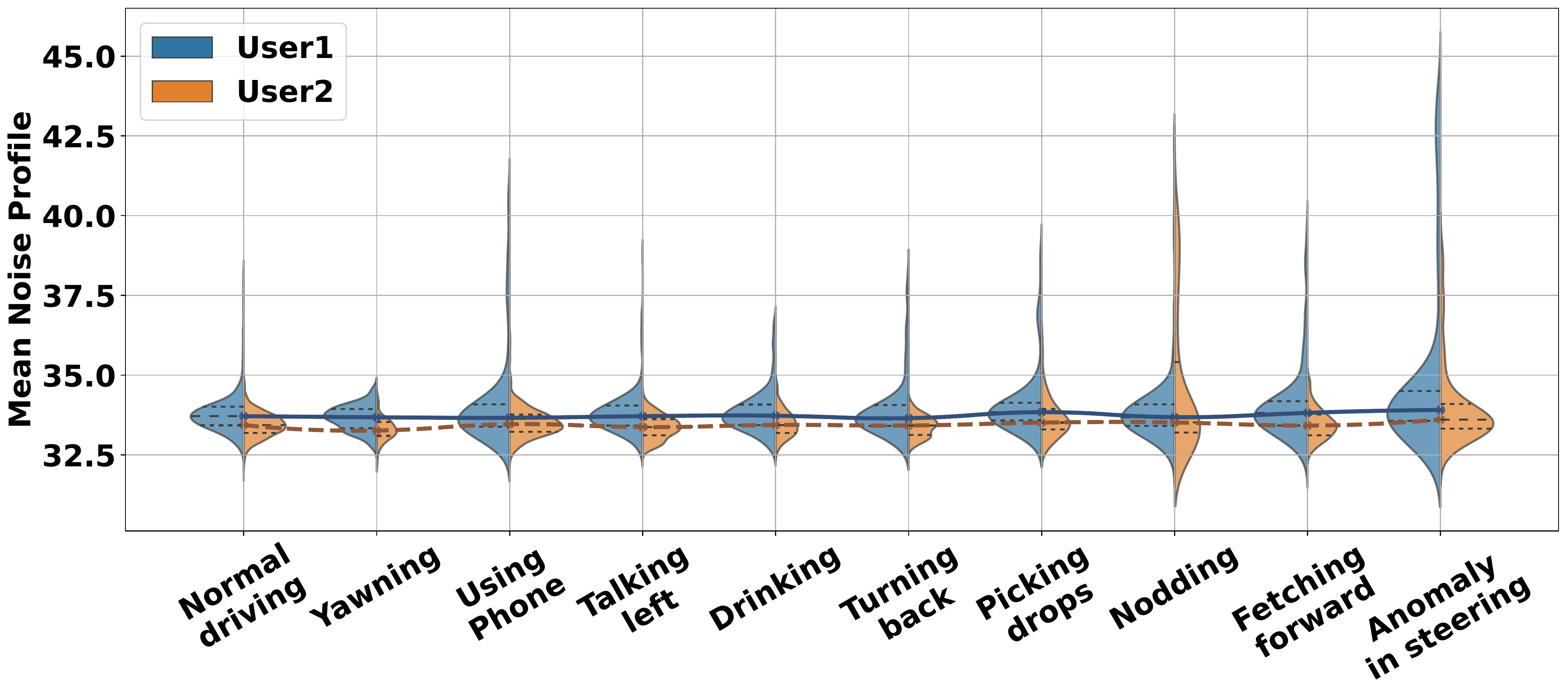}
            \label{fig:noise_violin}}
    \subfigure[]{
    \includegraphics[width=0.31\textwidth]{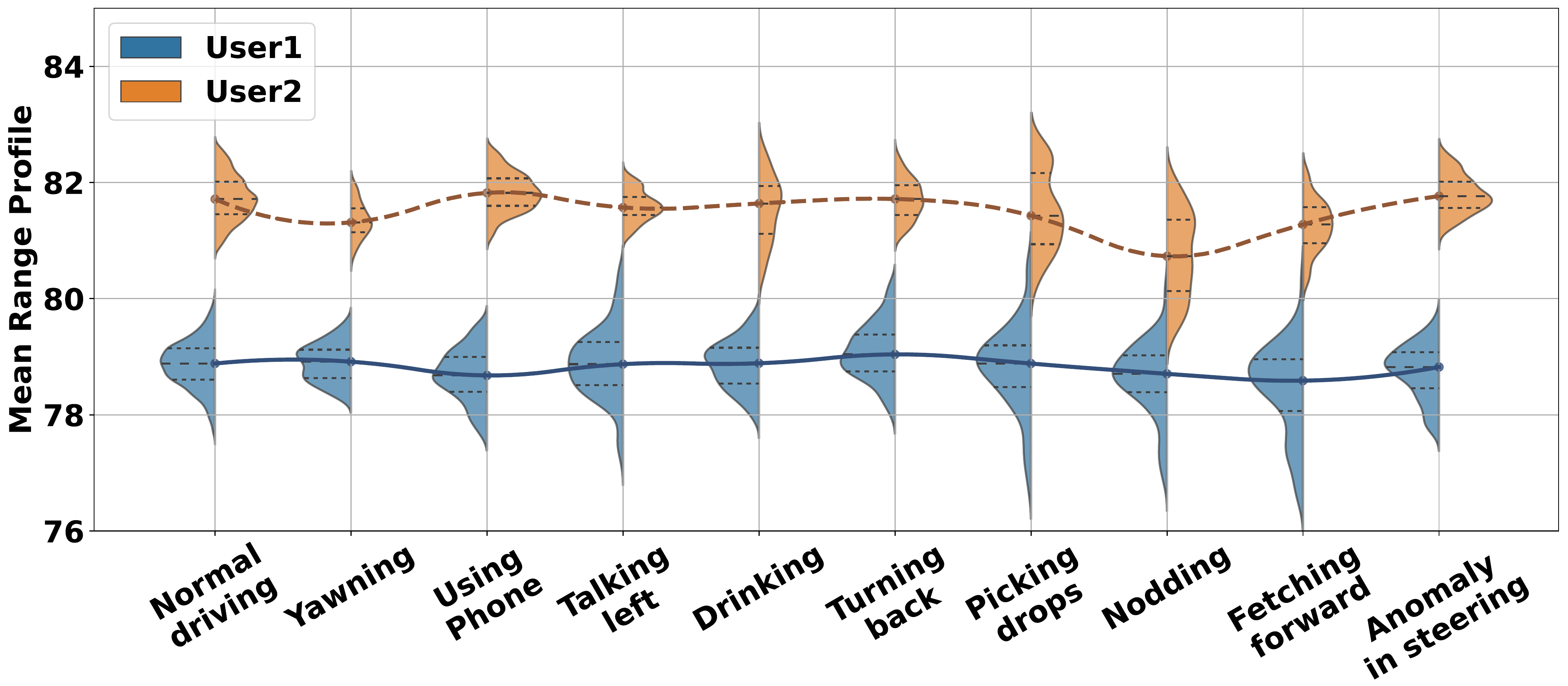}
            \label{fig:range_violin}}
    \caption{Variation in the mean of (a) range-doppler, (b) noise profiles, and (c) range profiles across different driving actions for two users}
    \label{fig:violin_variation}
\end{figure*}

\subsubsection{Analysis of range-doppler heatmap}
The \textit{range-doppler} information is in the form of a 2D image, where its \textit{abscissa is the range (marked by $32$ indices of the $64$ available, as per the specifications of \texttt{AWR6843ISK})}, the \textit{ordinate is the Doppler speed}, and the value contained is the magnitude of power for the velocity across the different range and doppler bins. The typical distance between the driver and mmWave sensor in the deployed setup is around $0.6$ to $0.8$ meters, corresponding to range bins from $6$ to $16$. Thus we observe a higher power value within this range, as seen in Fig.~\ref{fig:raw_range_doppler}. For body movements such as fetching forward, picking drops, nodding, turning back, etc., the driver moves their body in a particular direction, stays in that orientation for a few seconds, and then returns to normal position. For these types of actions, we observe that when the driver starts the activity, the doppler shifts up in the positive direction (or in the negative, depending upon the direction of the movement). When the driver returns to the normal state, the doppler shifts down (or up, depending on the movement order) and finally returns to zero as the activity ends. Thus, we observe a circular pattern in the range-doppler heatmap for the given actions, as shown in Fig.~\ref{fig:raw_range_doppler}. In the event of a steering anomaly, the driver abruptly moves the steering wheel, resulting in a sudden change in the range-doppler heatmap across different range bins. However, a driver usually makes fewer body movements for activities such as yawning, drinking, or using a mobile phone while driving. Thus the variation in the range-doppler shows lesser strength in the magnitude of the power value.

\subsubsection{Analysis of range-doppler, range profile, and noise profile for different drivers}
To understand the observed variations in more detail, we show the distribution of the mean of the range-doppler, the range profile, and the noise profile for different driving actions for two different drivers in Fig.~\ref{fig:violin_variation}. As can be seen from Fig.~\ref{fig:doppler_violin} and Fig.~\ref{fig:noise_violin}, yawning, using a phone, talking to the person in left (adjacent seat) or drinking shows lesser variation as the distribution is denser in the interquartile range. However, for the rest of the driving actions, we observe that the distribution is less dense towards the median representing some \textit{macro body movements} of the driver. Fig.~\ref{fig:range_violin} shows the distribution of range profile across different driving actions, where the variation is less significant across activities. However, the location of the medians of the range profile for the two users has a substantial gap due to differences in the driver's height, sitting stance, and position. 

\begin{figure}[!ht]
    \centering
     \includegraphics[width=0.4\textwidth]{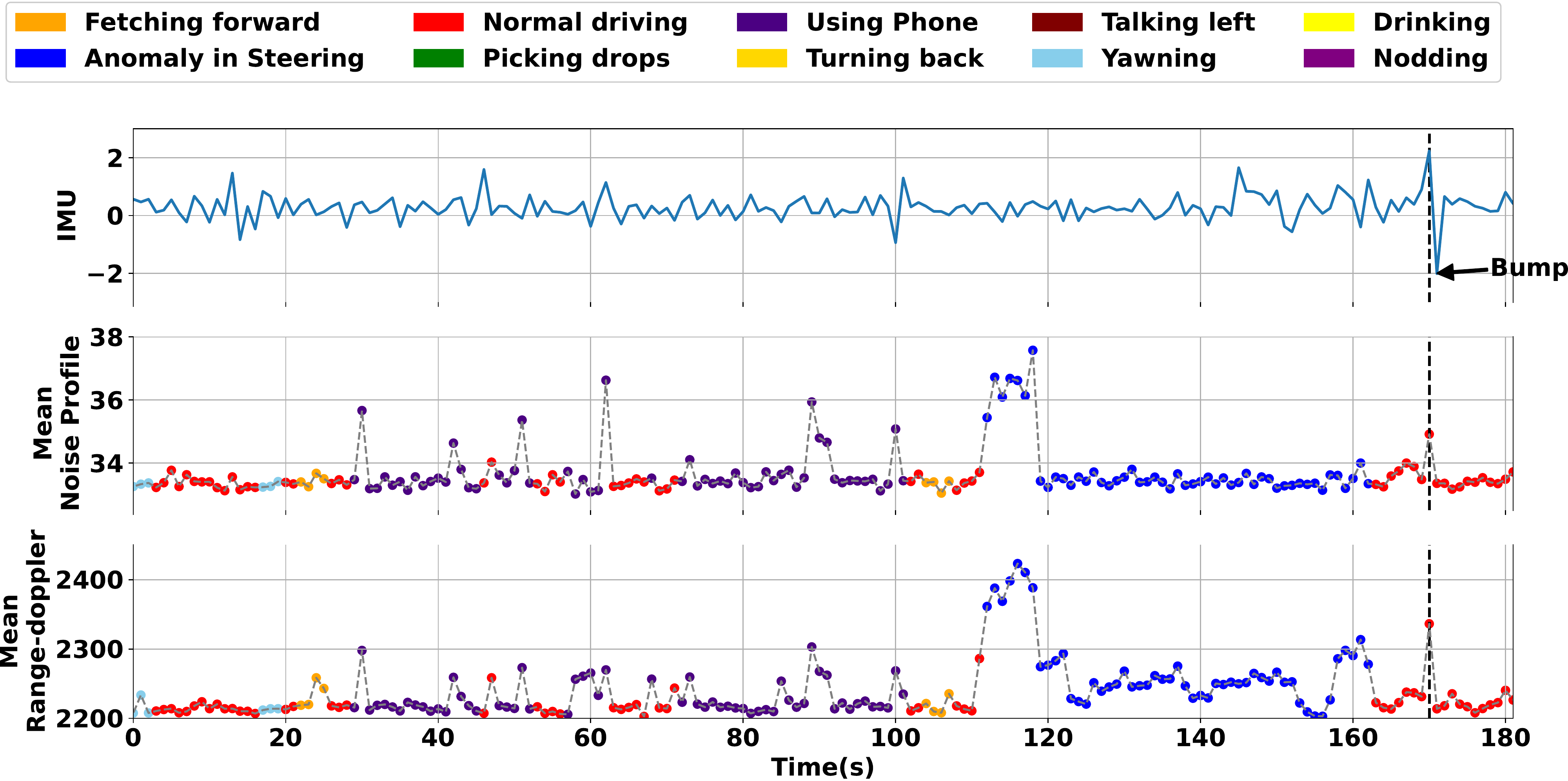}
    \caption{Variation in the IMU data and Radar measurements}
    \label{fig:mean_variation}
\end{figure}
\subsubsection{Analysis of different movements over time}\label{sec:obs}
Next, we study the impact of different movements and the corresponding signatures over time. The signs of our interest also include jerks caused by road bumps, which are captured by leveraging an IMU sensor. During road bumps, the z-axis of the IMU's accelerometer shows a peak signifying the magnitude of the movement caused. We plot the mean of range-doppler, noise profile, and the mean of the IMU data for each second over three minutes in Fig.~\ref{fig:mean_variation}. The figure shows that each activity has its period and mean value. For example, yawning takes place for a shorter duration and marks a low mean value (close to the mean of normal driving) of range-doppler. The driver uses his phone for a longer term, similar to normal driving behavior, except for certain peaks where the driver is making major body movements to pick up the phone or move his hand from the steering. On the other hand, an anomaly in steering takes place for a longer duration, with significant variation in the mean compared to normal driving. From this, we can conclude that each of these driving behaviors has its own signature in the range-doppler or noise profile and has its own time to completion. If we switch to the IMU information, there is no significant variation while the driver performs these activities. However, the IMU data shows a peak when there is a road bump (at around 170th second in Fig.~\ref{fig:mean_variation}). This peak is also observed in the mean of the range-doppler and noise profile data. Therefore, road bumps can directly impact the mmWave measurements.

These observations establish the fact that each of the driving behaviors has its own spatial as well as temporal variation. Therefore, for classifying the driving behaviors, we need to consider both \textit{spatial and temporal variations} in the feature space. At the same time, the instances where road bumps are involved should be handled adequately.
\begin{figure}[!ht]
    \centering
    \includegraphics[width=0.5\textwidth]{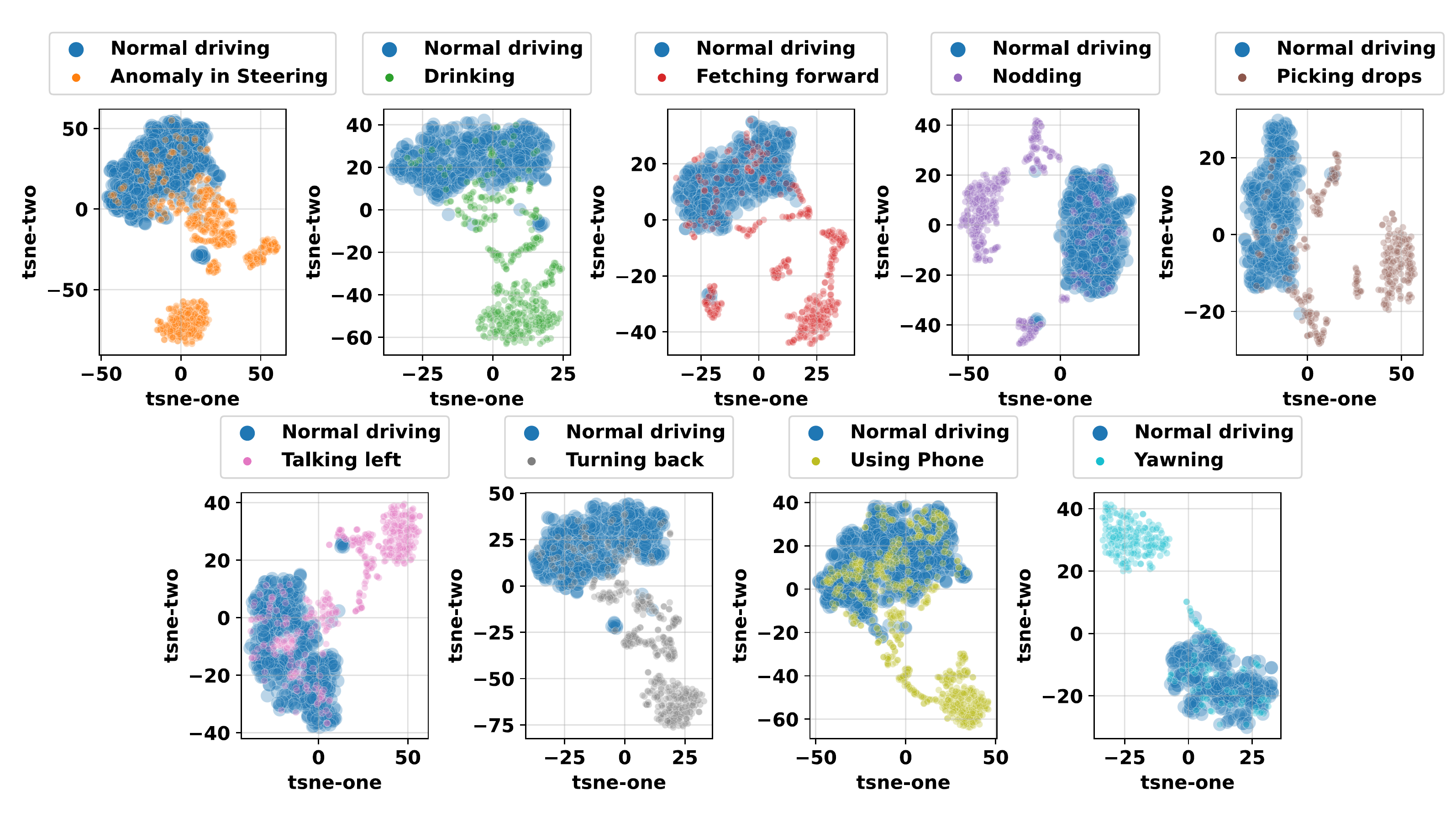}
    \caption{Distributions of normal driving with other dangerous driving actions in two-dimensional feature space}
    \label{fig:tsne_pca50_features}
\end{figure}

\subsubsection{Analysis of driving behavior distribution} Observing the mean of range-doppler, range profile, and noise profile, one can assess the feasibility of driving behavior classification. 
The range-doppler heatmap is a $16\times64$ 2D matrix representing $64$ range bins and $16$ doppler bins. 
On the other hand, the range profile and the noise profile are in the form of a 1D array of size $64$, representing $64$ range bins. Therefore, a total of $(16\times 64) + (64\times 2) = 1152$ dimensions are involved. We use a T-distributed neighbor embedding (t-SNE) based dimensionality reduction technique to lower the higher dimension ($1152$) space to a 2D space to visualize how each dangerous driving action is separated from the normal driving behavior. From the resultant Fig.~\ref{fig:tsne_pca50_features}, we observe that dangerous driving actions are somewhat separable from normal driving behavior. Nevertheless, no significant difference can be observed in the case of activity pairs such as normal driving and using the phone, normal driving and talking to the passenger, etc. Interestingly, it is worth noting that road bumps or any arbitrary sudden movements of the driver can also lead to variation in the range-doppler or noise profile.

The key takeaways from this analysis can be summarized as follows. (1) Range-doppler, noise profile, and range profile show unique patterns across different driving actions and drivers. (2) Each driving action has its unique signature at completion time. (3) Features of dangerous driving actions having micro-body movements show close signatures to normal driving behavior. (4) In real driving, road bumps can also lead to variation in the feature space and thus can confuse the model. Building on these takeaways, we next design \ourmethod for classifying different driving behaviors.

\section{Methodology}
\begin{figure}[!ht]
    \centering
    \includegraphics[width=0.8\columnwidth]{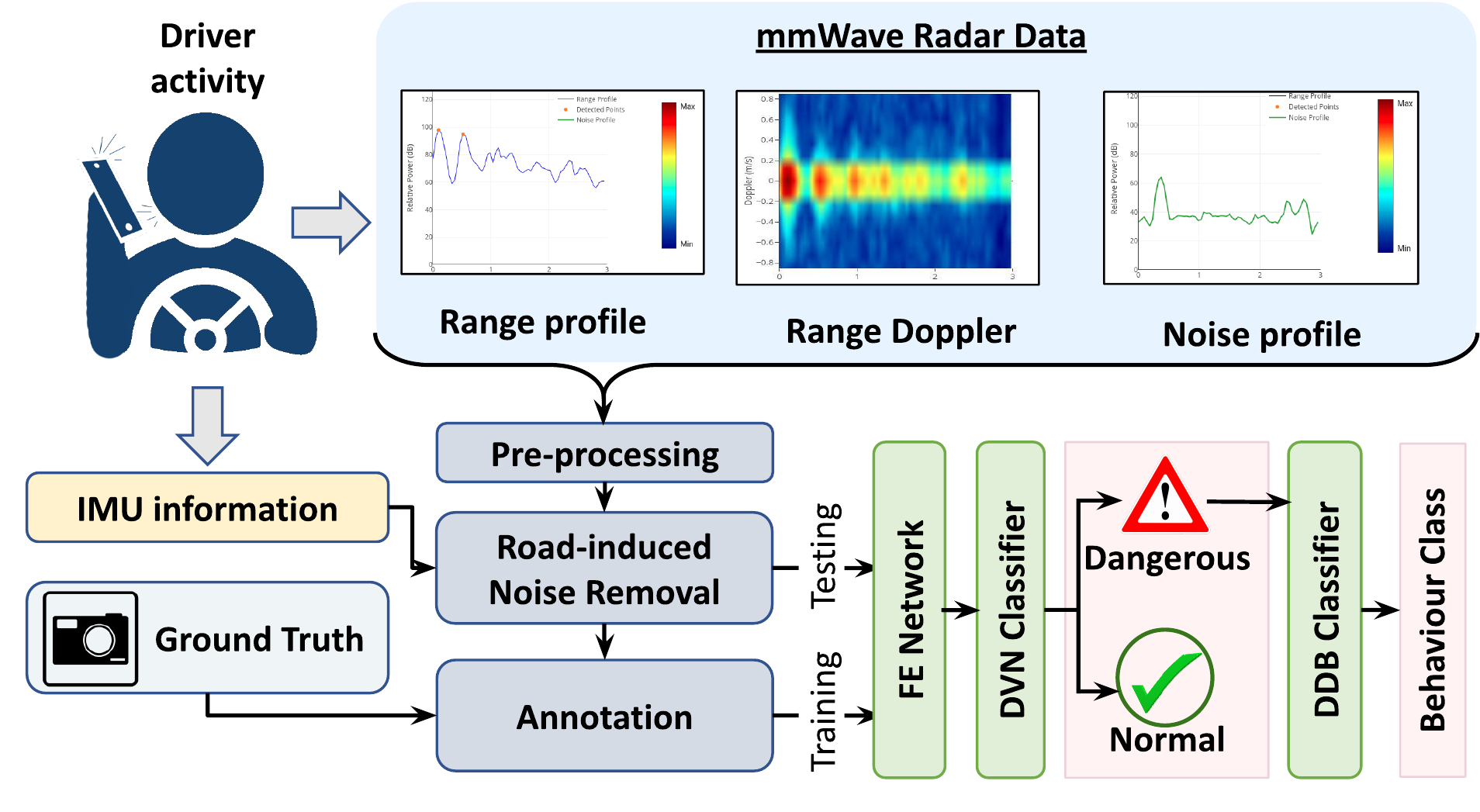}
    \caption{Overview of \ourmethod}
    \label{fig:overview}
\end{figure}

A broad overview of the processing steps involved in the formulation of \ourmethod is shown in Fig.~\ref{fig:overview}. An FMCW radar yields measurements corresponding to the driver's body movements. In contrast, an IMU sensor captures vehicle-specific mobility patterns. This information is fed to \ourmethod for further processing. We describe the subsequent steps as follows.

\subsection{Pre-processing}
As discussed in Sec.~\ref{sec:back}, primary features involved in sensing driving behaviors include range-profile, noise-profile, and range-doppler heatmap. Notably, individual driving actions are not a single instantaneous activity; instead, they occur for a while and have a unique temporal pattern in the associated features. In order to capture this temporal variation, we concatenate $10$ feature frames together, determined based on empirical analysis as discussed in Sec.~\ref{sec:eval}. Finally, we apply a \textit{min-max scaler} to normalize the features from 0 to 1.

\subsection{Removal of Road-induced Noise}
\label{sec:noise_rem}
As observed in Sec.~\ref{sec:back}, poor road conditions (i.e., broken roads, speed bumps, etc.) can directly impact the noise profile and range-doppler heatmap. This variation in the feature space is similar to the impact due to the driver's driving actions, thus introducing ambiguity if not appropriately handled. In order to detect such unwanted variations, we take the help of an alternate modality from IMU sensors embedded within the device. Following the existing works~\cite{mandal2021exploiting,xu2017er}, we use the z-axis acceleration of the vehicle to determine jerkiness due to poor road conditions or bumps. Subsequently, \ourmethod filters the corresponding data collected from the mmWave sensor to suppress unwanted noise in the feature space.

\subsection{Classification Pipeline}
\begin{figure}
    \centering
    \includegraphics[width=0.35\textwidth]{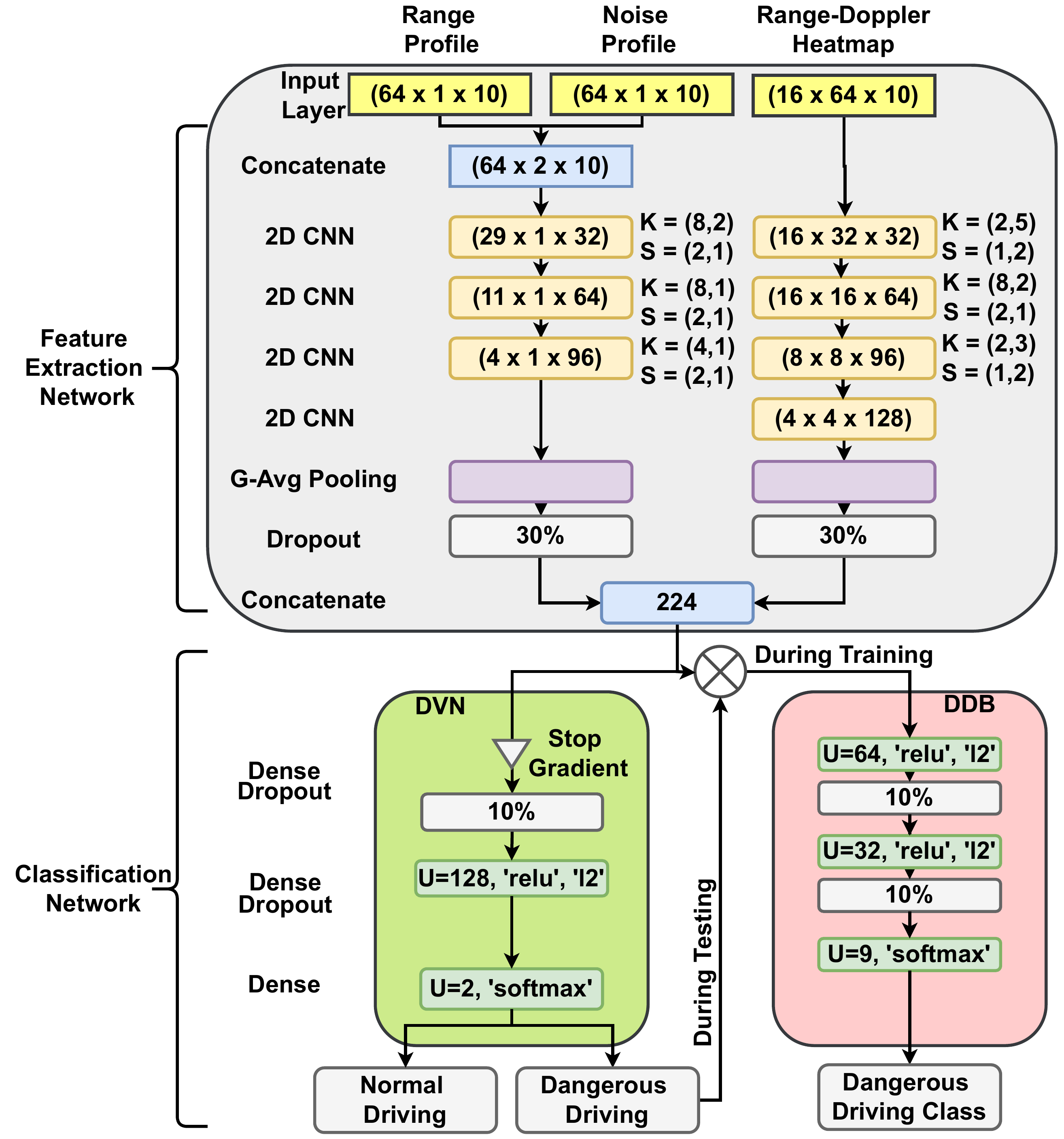}
    \caption{Fused-CNN Model Architecture}
    \label{fig:model}
\end{figure}

Unlike works based on traditional Convolutional Neural Networks (CNN) and range-doppler spectrogram~\cite{wang2016interacting,liu2021m, tiwari2021mmwave}, we take advantage of range-doppler as well as range profile and noise profile features for classifying normal and dangerous driving behaviors. We propose a novel Fused-CNN architecture as shown in Fig.~\ref{fig:model}; the model details follow. 

\subsubsection{\textbf{F}eature \textbf{E}xtraction (\textbf{FE}) Network}
The range profile provides maximum power value at zero-doppler regions. On the other hand, the noise profile provides noise floor power at non-zero Doppler regions. Thus concatenating such information helps in understanding the spatial dynamics of the driver across the overall range bins. We concatenated $10$ feature frames, as discussed earlier, to capture the temporal variations. Hence, range and noise profiles are vectors of size $64\times1\times10$, and concatenating them together forms an array of size $64 \times2\times 10$. On the other hand, the range-doppler heatmap forms a stacked 2D image-like feature with size $16\times64\times10$.

We use three Convolutional 2D layers with valid padding and ReLU activation, followed by a global average pooling to extract $96$ dimensional range-noise-based embeddings. Like range-noise feature extraction, we apply convolution 2D operation to extract the dependency of neighboring values within all possible $k\times k$ regions at each range-doppler frame along with the temporal relationship of past $10$ such frames. Further, it computes several cross-channel feature maps, which are helpful for subsequent model layers. We use four 2D Convolutional layers with the same padding and ReLU activation, followed by a global averaging to extract $128$ dimensional range-doppler feature embeddings. Next, the concatenated range-noise and range-doppler feature embeddings are forwarded through two successive modules of the proposed architecture.

\subsubsection{\textbf{D}angerous \textbf{D}riving \textbf{B}ehaviors (\textbf{DDB}) Classifier}
The DDB classifier takes the cross-channel feature embeddings from the FE Network and passes through two successive Dropout and Dense layers, where the dropout rate is kept as $10\%$ to prevent overfitting. The last layer has $9$ neurons with softmax activation to output a joint probability distribution over the nine dangerous driving actions. Note that the FE Network is trained only with the backward gradients from the DDB Classifier to learn efficient feature extraction for classifying $9$ dangerous driving behavior.

\subsubsection{\textbf{D}angerous \textbf{V}s \textbf{N}ormal driving (\textbf{DVN}) Classifier}
The DVN Classifier utilizes the learned feature embeddings from the pre-trained FE Network to classify all forms of dangerous driving behaviors from normal driving. The weights learned in the FE Network during the training of the DDB Classifier are unaltered due to backward gradient stopping beyond DVN Classifier. Such an approach helps optimize the net training time and keeps the model lightweight and feasible for real-time deployment. 

\subsubsection{Lazy Inferencing} During real-world deployment, the DVN Classifier determines whether the driving behavior is normal or dangerous. Upon detecting normal driving behavior, \ourmethod differs the execution of the DDB Classifier, reducing computational overhead and energy consumption. The DDB Classifier is queried only if the DVN Classifier detects potential dangerous driving behavior, further determining the particular dangerous driving class.

\section{Implementation and Evaluation}
We deploy the experimental setup in real driving scenarios to study the practical implications of dangerous driving behaviors. The experiment is carried out across $3$ cars and $5$ drivers. The radar is placed on the dashboard of the car. On average, the distance between the driver and the radar is around $0.7$ meter. After thorough data collection and careful ground truth annotation, we evaluate our proposed Fused-CNN model against two state-of-the-art baselines. The detail follows. 

\subsection{Hardware Setup} \ourmethod is implemented on COTS mmWave radar \texttt{AWR6843ISK}, an FMCW-based mmWave radar from Texas Instruments. The radar works in the frequency range of $60$-$64$GHz with a range resolution of approximately $4$cm, which is adequate for measuring the activities of interest listed in Sec.~\ref{sec:pilot_study}. The radar's maximum range is up to a distance of $10$ meters with a Field-of-View of $-70\degree$  to $+70\degree$ on azimuthal and elevation planes. This Field-of-View is suitable for detecting dangerous driving behaviors within the car. \ourmethod measures variations across $64$ range bins, which represent $2.4$ meters distance from the dashboard. This brings a range resolution of $3.75$ cm. \ourmethod collects the doppler information across $16$ doppler bins from the 2D-FFT to have a velocity resolution of $0.13$ m/s which is sufficient to capture the driver's micro as well as macro body movements in real-world driving scenarios. \tablename~\ref{tab:radar_conf} shows various configuration parameter of the mmWave radar. Moreover, we have used Raspberry Pi 4 Model-B with 8GB RAM for live driving behavior detection. The mmWave radar and IMU sensor are connected to the Pi-4 via USB and I2C bus, respectively. The IMU sensor filters out the road-induced noises as discussed in Sec.~\ref{sec:noise_rem}. The mmWave serial data is parsed and forwarded to the classification pipeline for inferring the driving behavior.

\begin{figure}
    \centering
    \subfigure[]{
    \includegraphics[width=0.45\columnwidth]{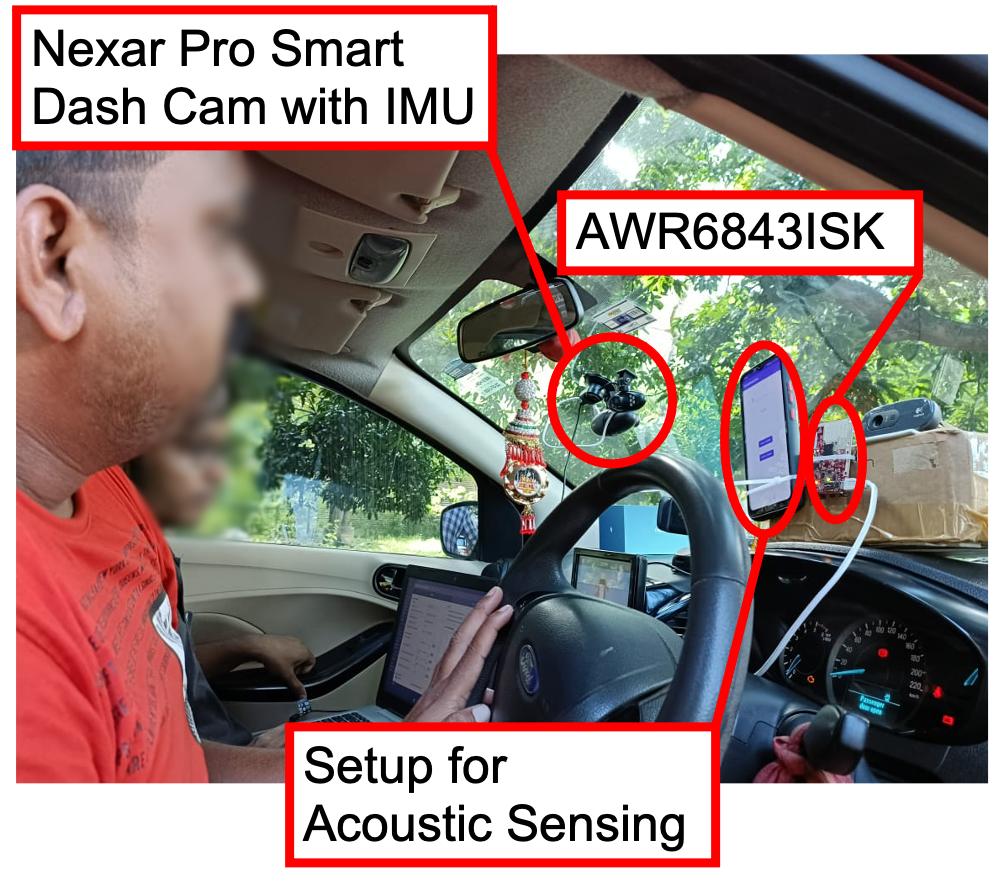}
    \label{fig:hardwaresetup}
    }
    \hspace{-10pt}
    \subfigure[]{
    \includegraphics[width=0.23\textwidth]{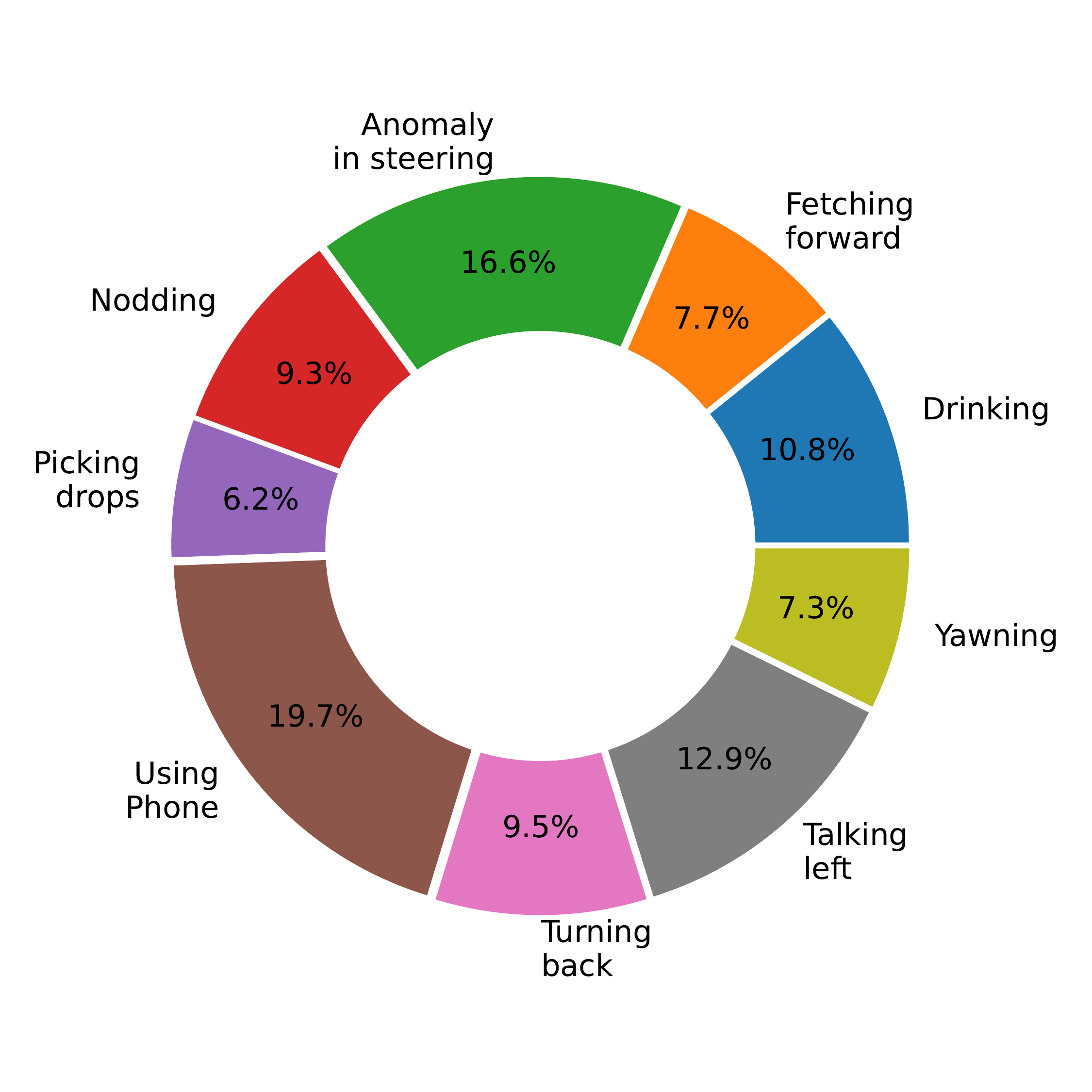}
    \label{fig:data_dist}}
    \caption{(a) Hardware setup (b) Collected data distribution over different dangerous driving behaviors}
\end{figure}

In order to collect the ground truth of the driving actions, we have used Nexar Pro Smart Dash Camera\footnote{\url{https://www.getnexar.com/global/the-dash-cams}}. The collected videos from the Nexar are used to annotate the mmWave Radar-generated features. \figurename~\ref{fig:hardwaresetup} depicts the hardware setup used for field experimentation. We provide more details on the data collection strategy in the subsequent subsection.

\begin{table}[]
\scriptsize
\centering
\caption{Radar Configuration}
\label{tab:radar_conf}
\begin{tabular}{@{}ll@{}}
\toprule
\textbf{Parameters}              & \textbf{Value}           \\ \midrule
Start Frequency (GHz)            & $60$                       \\
Range Resolution (m)             & $0.0375$                   \\
Maximum Unambiguous Range (m)    & $2.41$                     \\
Maximum Radial Velocity (m/s)    & $1$                        \\
Radial Velocity Resolution (m/s) & $0.13$                     \\
Frames per Second                & $5$                        \\
Number of chirps per frame       & $64$                       \\
Baud-rate       & $921600$ bps   
\\
\bottomrule
\end{tabular}
\end{table}

\subsection{Data Collection}
To collect raw mmWave features with the hardware mentioned above, we have used mmWave-Demo-Visualizer~\footnote{\url{https://dev.ti.com/gallery/view/mmwave/mmWave_Demo_Visualizer/}} and modified the source code to enable raw data collection with users input. Data collection is carried out for $5$ users for a total duration of $20$ hours ($4$ hours for each user), using three different vehicles, of which two were sedans, and one was an SUV. Among the drivers, one was female, and the rest were male, with ages ranging from $27$ to $45$. The least experienced driver has one year of driving experience, while the most experienced driver has more than ten years of driving experience. Collected data contains mmWave radar data (i.e., range-profile, noise-profile, and range-doppler heatmap) along with IMU, GPS, video, and audio of the roadside, as well as the interior view of the car's cabin collected using Nexar. We took help from three volunteers to annotate dangerous driving behaviors from the videos captured by the interior car camera.

\subsection{Software Setup and Baselines}\label{sec_baseline}
We have used \texttt{Python 3.9.6}, \texttt{Tensorflow v2.10.0}, and \texttt{Scikit-learn v1.1.2} for implementing the Fused-CNN-based driver behavior classifier model alongside other two state-of-the-art baselines: Random Forest (RF), and VGG-16~\cite{simonyan2014very}. The models are trained on an iMac (with 16GB primary memory running MacOS v12.6 with base-kernel version: 21.6.0) and for the hyper-parameter tuning we utilized a Workstation (48 $\times$ vCPU, GPU Nvidia TitanX 12GB, \& RAM 128GB).

The RF model's features are engineered to take min, max, mean and standard deviation of the range profile, noise profile and range-doppler within a kernel size of $16\times1$, $16\times1$ and $16\times16$ for the total $10$ frames with an array size of $64\times1$, $64\times1$ and $16\times64$ respectively, resulting \begin{math}(\frac{(64\times1)}{(16\times1)}+\frac{(64\times1)}{(16\times1)}+\frac{(16\times64)}{(16\times16)})\times4\times10 = 480\end{math} features. We have taken the Random Search Cross Validation approach~\cite{koehrsen2018hyperparameter} to search for the best hyperparameters for the RF classifier within a wide range of values for each hyperparameter, performing K-Fold cross-validation with each combination of the hyperparameter values. Based on that, we have selected $400$ estimators.

On the other hand, we have used VGG-16~\cite{simonyan2014very} network as another baseline. This network is initialized with the pre-trained weights on the ImageNet~\cite{deng2009imagenet} dataset to have a transfer learning approach helping in learning feature extraction from high dimensional image-like input data. On top of this base VGG-16 model, we have added global average pooling and successive Dropout and Dense layers in the Fused-CNN architecture to classify dangerous driving behaviors. The models are trained with a train-test split of $70\%$-$30\%$ and a validation split of $20\%$ from the training set. 
 
Besides these two baselines, we use FMCW based-acoustic modality as our third baseline. Previous works such as~\cite{jiang2021driversonar,xie2019d,xie2020real} use an acoustic sensing approach to detect dangerous driving behaviors with fewer behavior classes. To understand the feasibility of this approach with nine different dangerous driving behaviors, we implement acoustic-FMCW using the smartphone-embedded microphone and speaker in the near-ultrasound range between 16-19kHz. We transmit and receive raw chirps using the smartphone and then apply Range-FFT in the post-processing to generate the amplitude and phase of the IF signal across different range bins. We further apply 2D-doppler FFT to generate the range-doppler heatmap. Finally, using a Random Forest-based classifier, we compare its performance with \ourmethod{}.

\subsection{Results}
\label{sec:eval}
We first discuss the impact of concatenating (stacking) the temporal feature frames measured from the mmWave radar on the detection accuracy and then discuss the detailed results based on the optimal frame stacking. 

\begin{figure}
    \centering
    \subfigure[]{
    \includegraphics[width=0.15\textwidth]{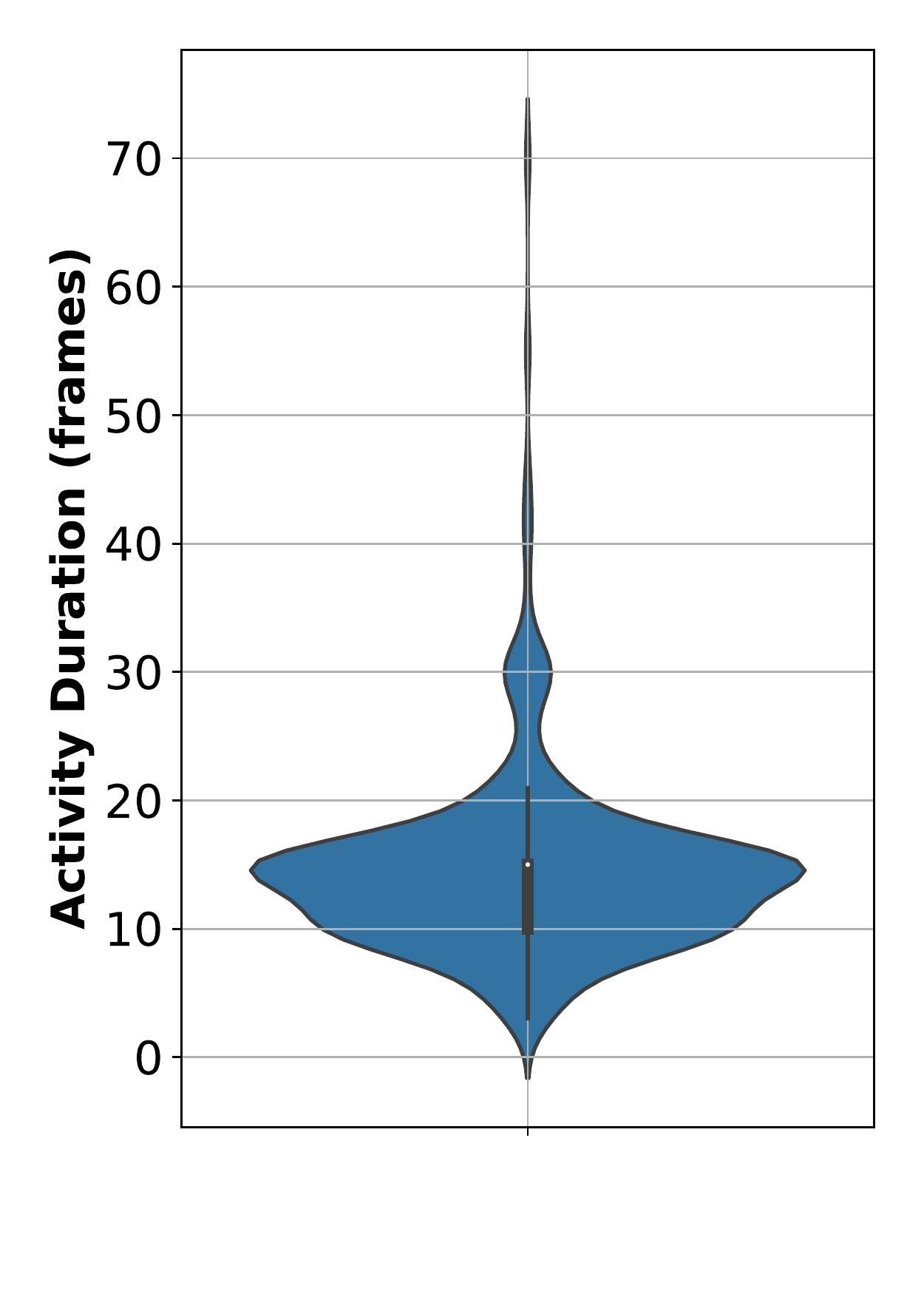}
    \label{fig:frame_dist}
    }
    \subfigure[]{
    \includegraphics[width=0.30\textwidth]{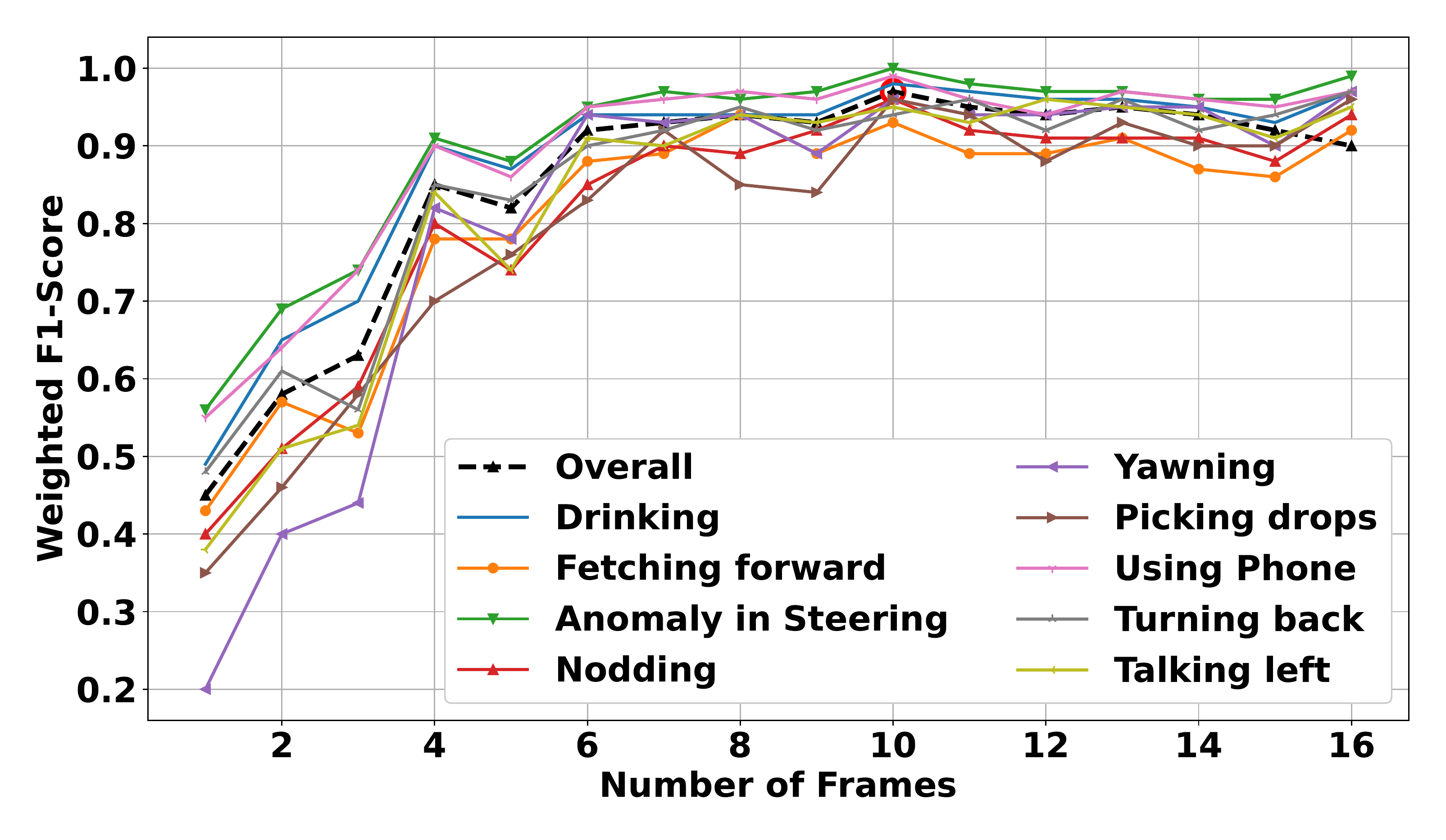}
    \label{fig:frame_perf}}
    \caption{(a) Distribution of Frames required to complete individual driving actions, (b) Variation in the weighted F1-Score with number of stacked frames}
    \label{fig:frame_stack}
\end{figure}

\begin{figure*}[t]%
	\centering
	\subfigure[]{%
		\label{fig:rf_cfm}%
		\includegraphics[width=0.24\textwidth]{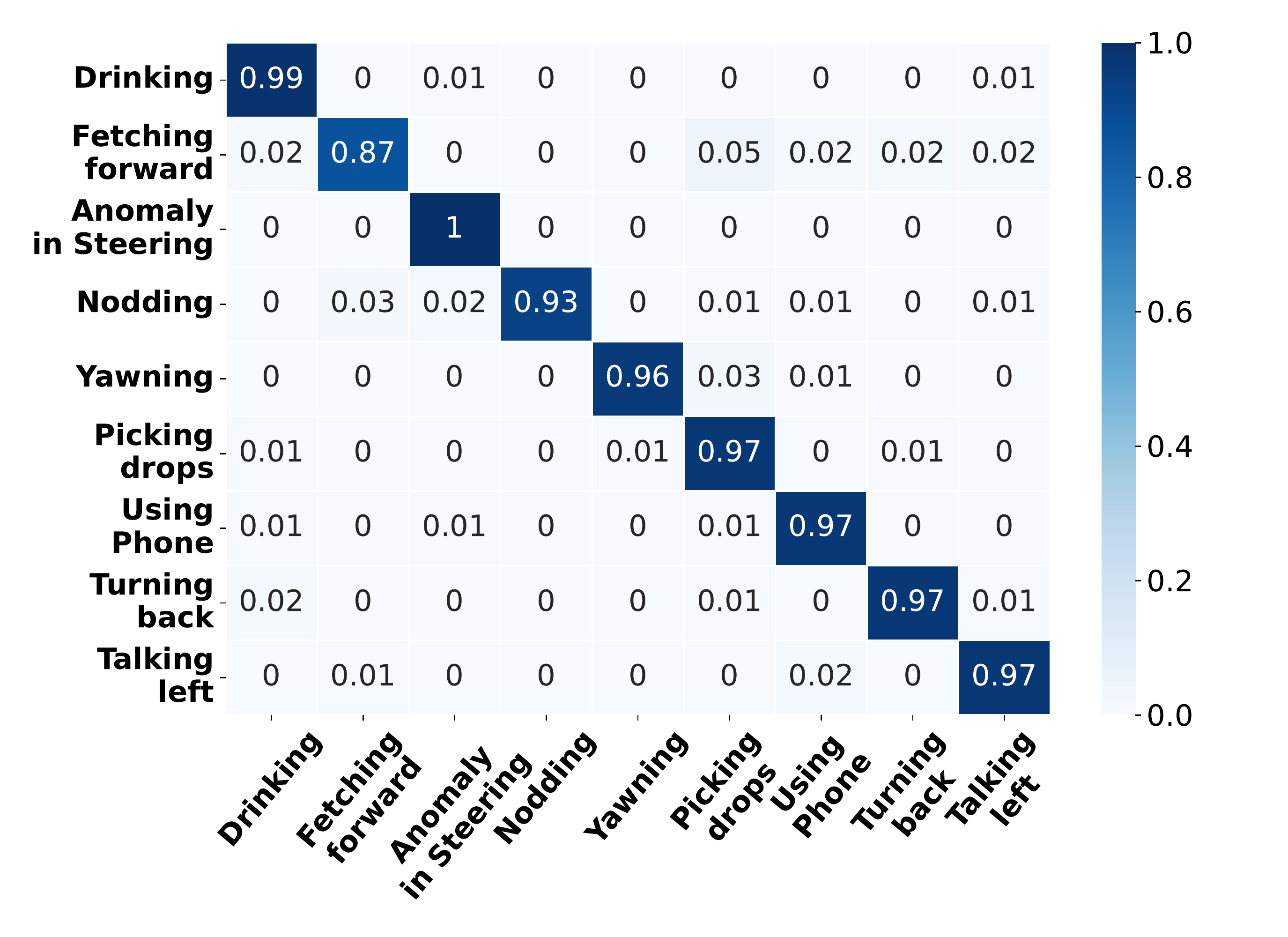}}\hfil%
	\subfigure[]{%
		\label{fig:static_cfm_cnn}%
		\includegraphics[width=0.24\textwidth]{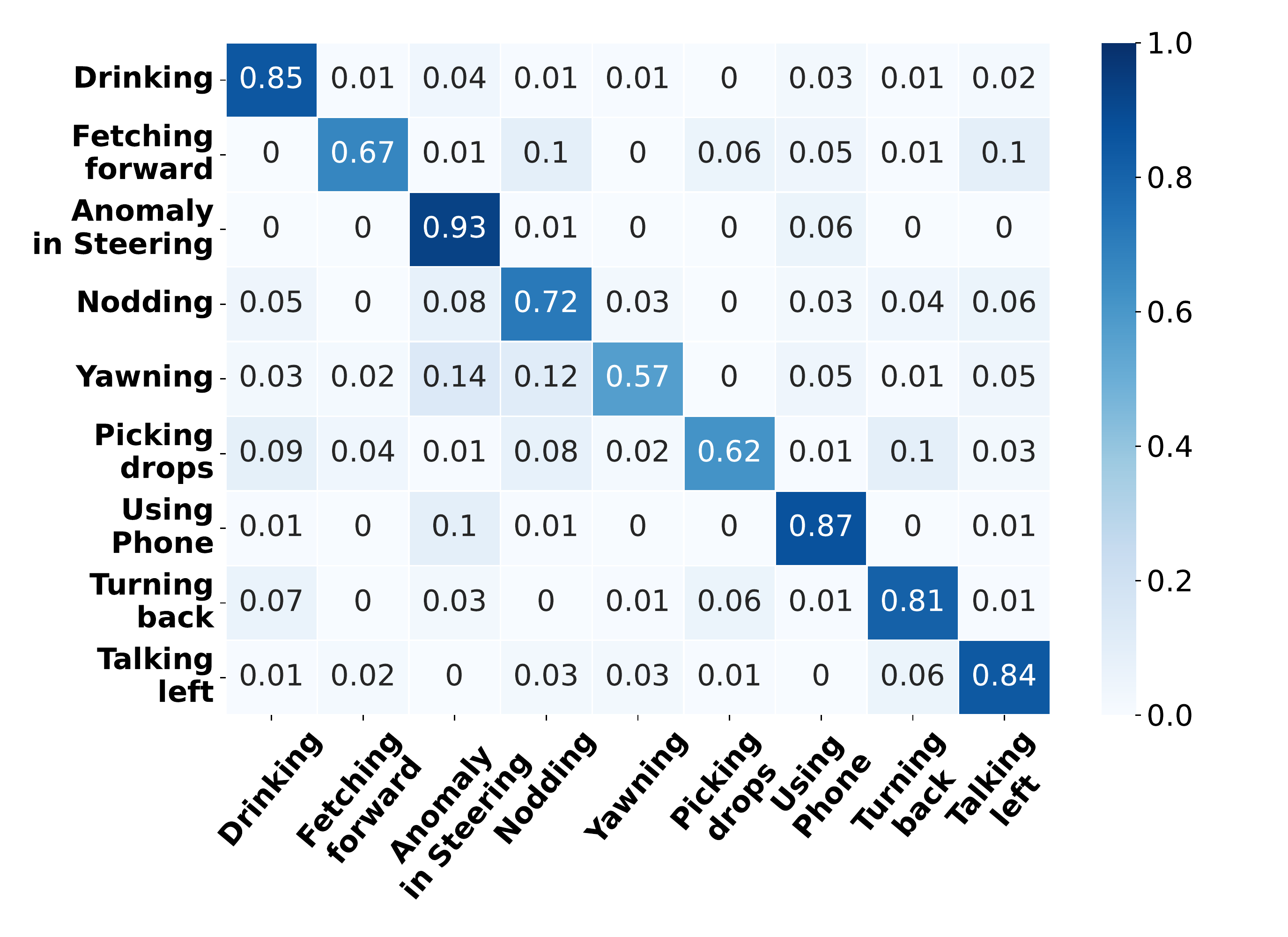}}\hfil%
	\subfigure[]{%
		\label{fig:static_cfm_vgg}%
		\includegraphics[width=0.24\textwidth]{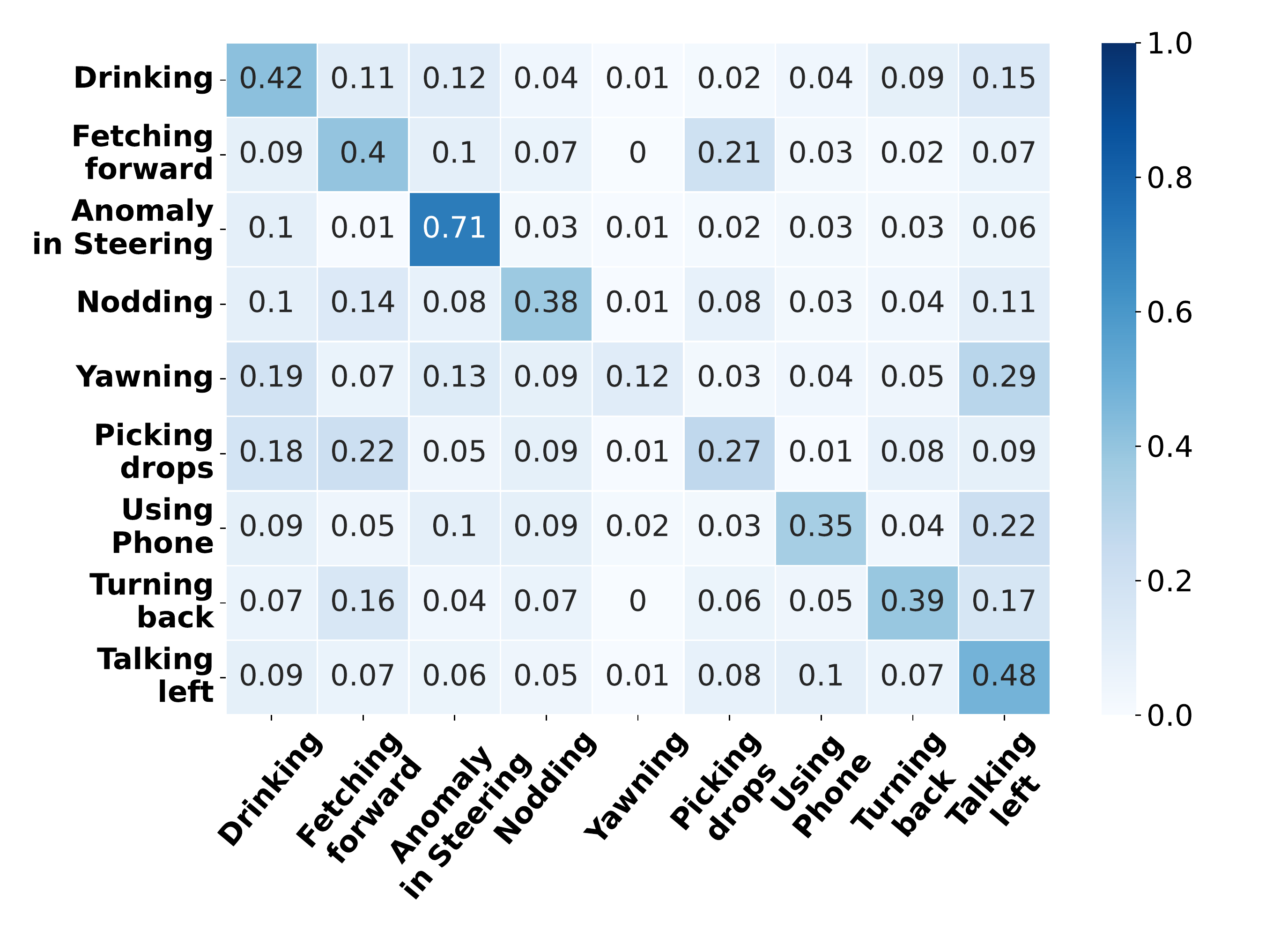}}\hfil%
	\subfigure[]{
	    \label{fig:acoustic_driving}
	    \includegraphics[width=0.24\textwidth]{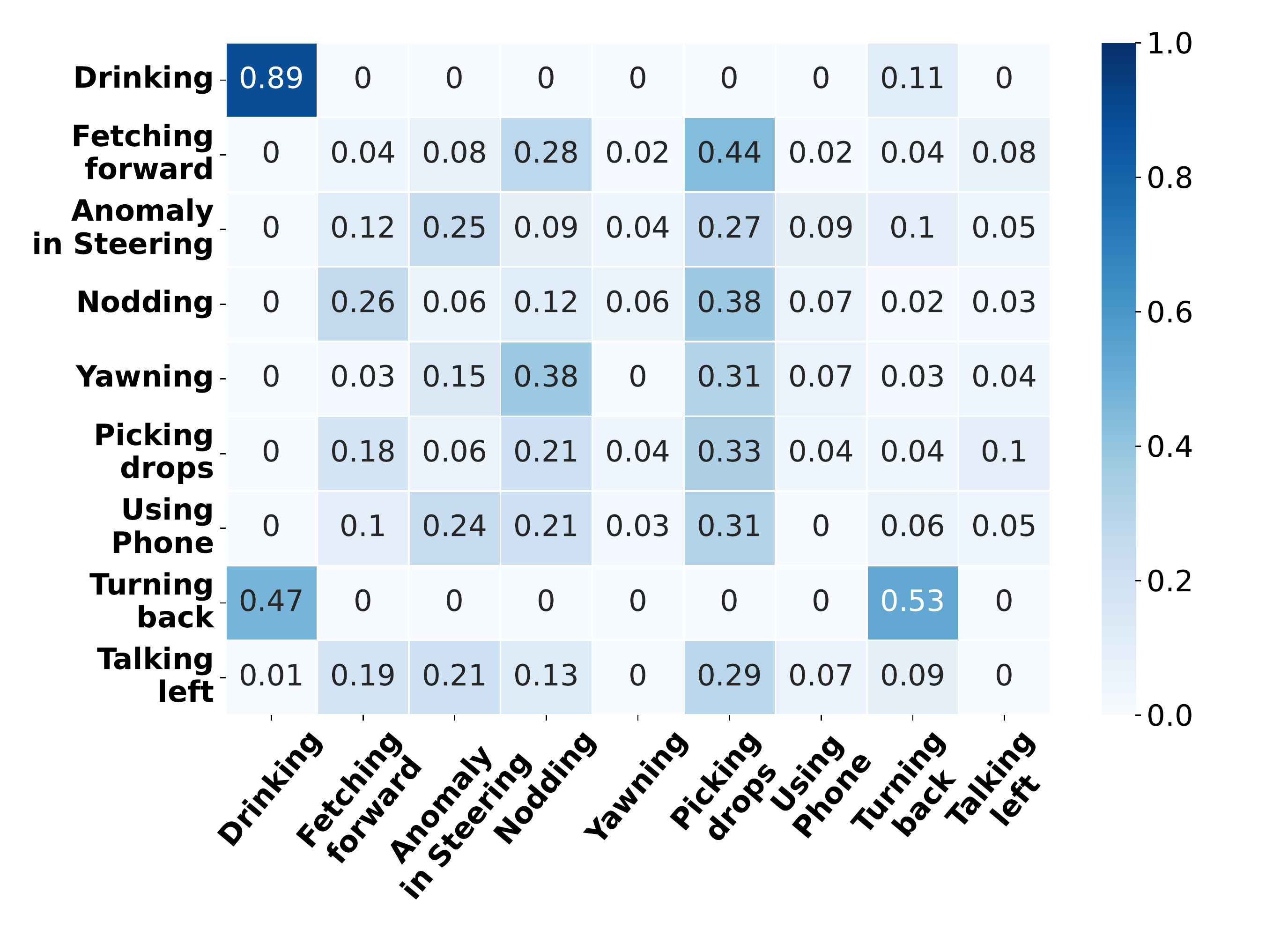}}\hfil%
	\caption{Confusion matrix for all the dangerous driving behaviors - (a) Fused-CNN, (b) RF, (c) VGG-16, (d) Acoustic-FMCW}
	\label{fig:driving_cfm}
\end{figure*}

\subsubsection{Impact of Frame Stacking on F1-Score}
As shown in \figurename~\ref{fig:frame_dist}, the time taken for each driving action is heavily skewed, and its median lies at around $16$ feature frames measured from the mmWave radar. Thus, driving actions have a temporal impact on the mmWave radar measurements. We stack multiple frames in order to capture this temporal impact. To determine the optimal frame stacking for detecting dangerous driving actions, we vary the number of frames from $1$ to $16$ as shown in \figurename~\ref{fig:frame_perf}. The figure shows that frame stacking has a direct impact on the overall performance of the classification pipeline. We observe that on increasing the number of frames, some activities that take longer to complete show a better F1-Score. For example, using a phone or anomaly in the steering takes a longer duration; thus, more frame stacking helps achieve better performance. However, for behaviors with shorter duration (i.e., fetching forward or nodding), F1-Score starts dropping. From \figurename~\ref{fig:frame_perf}, we see that \ourmethod{} achieves highest overall F1-Score for stacking 10 frames (activity time window of 2 seconds). Accordingly, we utilize $10$ stacked frames throughout this paper.

    

\begin{figure}
    \centering
    \subfigure[]{
    \includegraphics[width=0.23\textwidth]{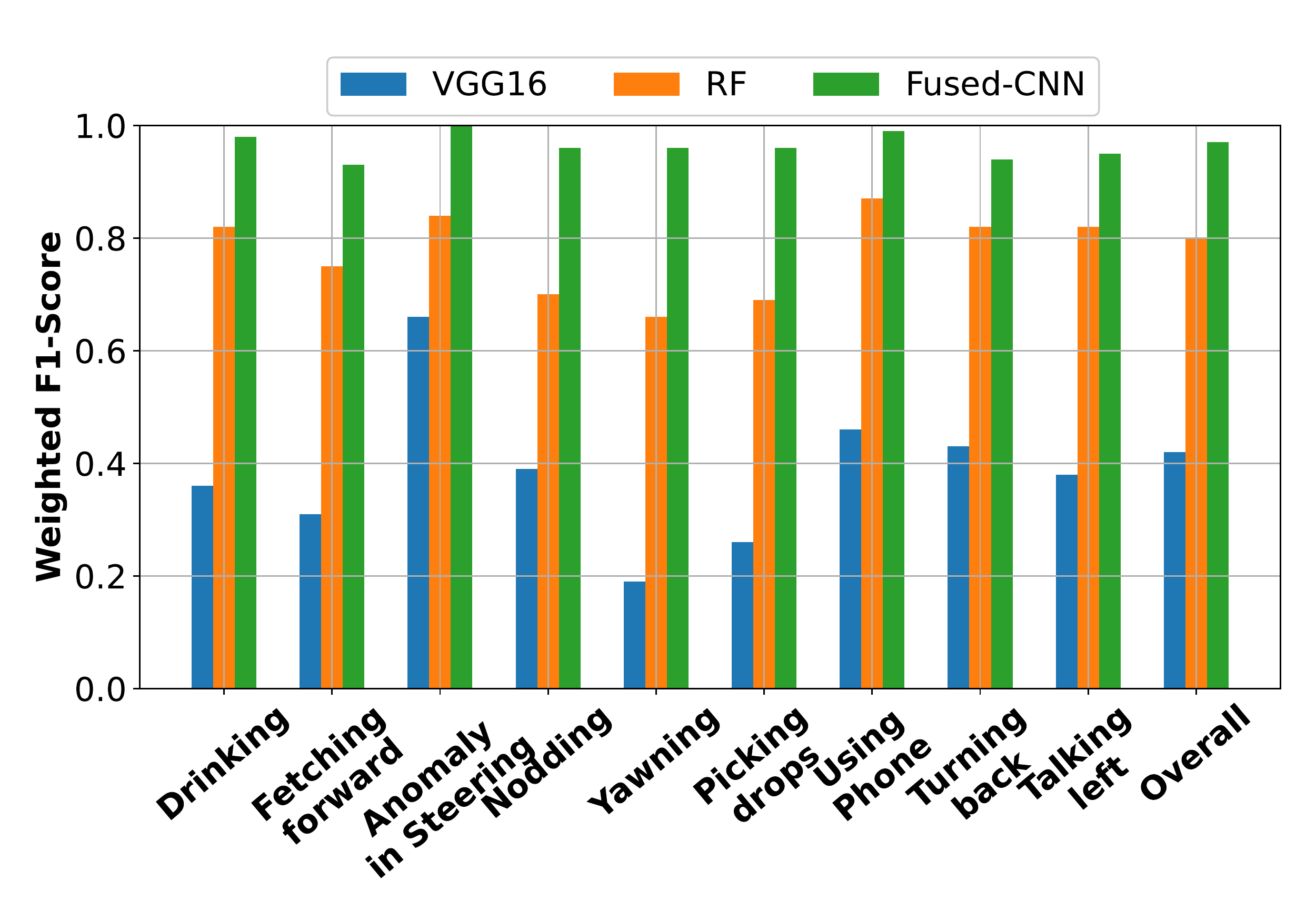}
    \label{fig:ddb_f1}}
    \subfigure[]{
    \includegraphics[width=0.21\textwidth]{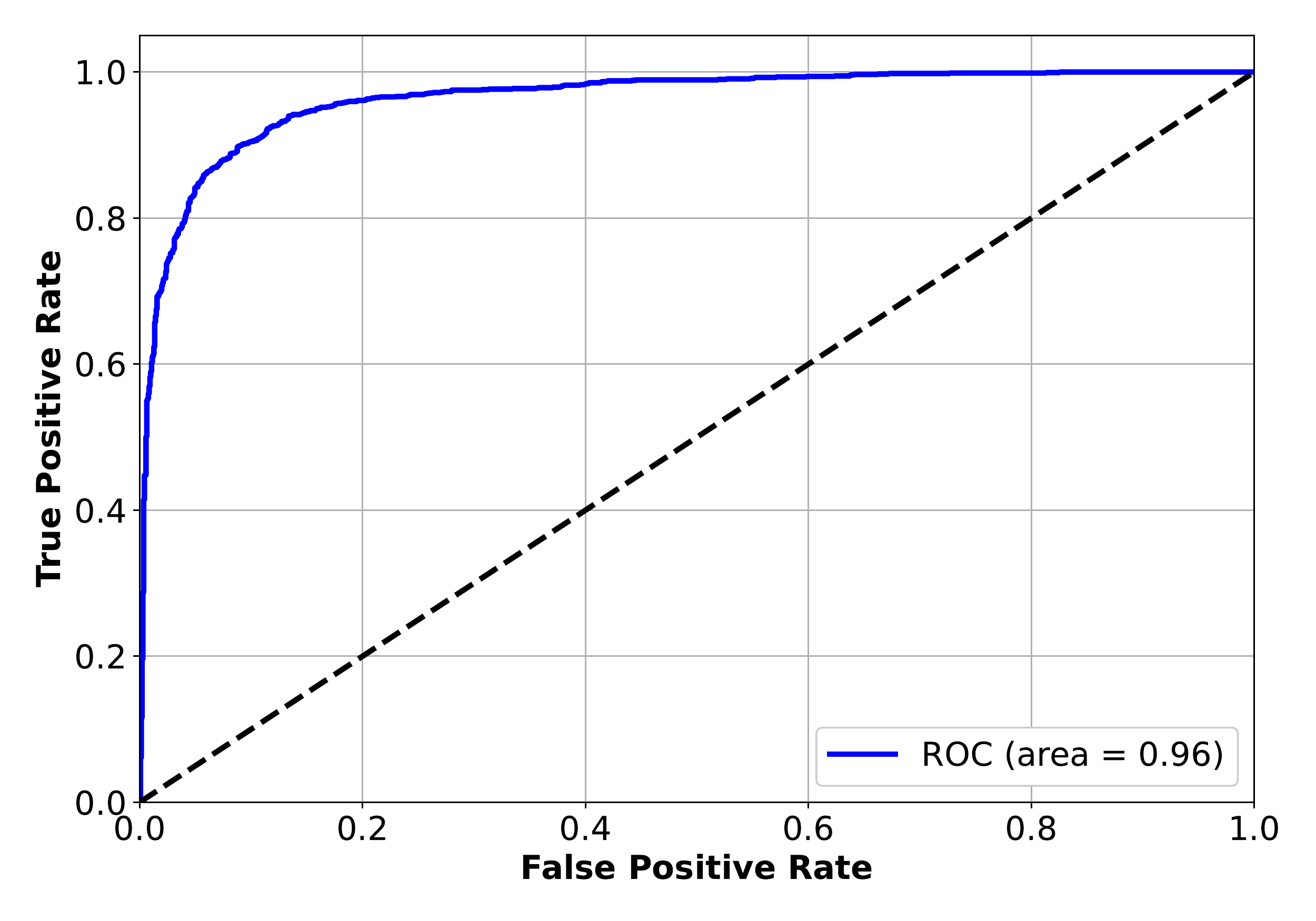}
    \label{fig:dvn_auc_roc}}
    \caption{(a) Weighted F1-Score across driving behaviors, (b) AUC-ROC for classifying dangerous vs normal driving.}
    \label{fig:overall}
\end{figure}

\subsubsection{Performance of DDB Classifier}
We compare the performance of the proposed Fused-CNN model with two baselines -- (i) Random Forest (RF) and (ii) VGG-16. In \figurename~\ref{fig:driving_cfm}, we report the confusion matrix for all the three classifiers. From the confusion matrix, it is evident that our proposed Fused-CNN model shows superior accuracy when compared with the baselines. \figurename~\ref{fig:ddb_f1} shows the average weighted F1-Score for all the individual dangerous driving actions. Among the baselines, RF performs better in comparison to VGG-16. The primary reason behind poor performance with VGG-16 is that this model expects a 2D input feature, which is a 2D range-doppler heatmap image in this case. Thus, it cannot take advantage of range or noise profile-based features. Also, VGG-16 is pre-trained with the \textit{imagenet} dataset and does not fit well in engineering features from a 2D range-doppler heatmap, which is mostly sparse across the range bins except for the range bin where the driver is present. In the case of RF, the features are passed across different kernel sizes to capture the spatial variation in the range and noise profiles as well as in the range-doppler heatmap. Thus we observe a slightly better performance in comparison to VGG-16. On the other hand, the proposed Fused-CNN has complete freedom in learning the spatio-temporal cross-features, as the features are not hand-engineered like RF (min, max, mean, standard deviation). Moreover, Fused-CNN shows a lower inference time due to less number of convolutional layers as compared to VGG-16.

\subsubsection{Performance of DVN Classifier}
In \figurename~\ref{fig:dvn_auc_roc}, we report the ROC curve for the DVN classifier. As shown in the figure, the area under the curve (AUC) is $0.96$, which ensures good accuracy in classifying dangerous driving from normal driving behavior. Moreover, we observe a weighted F1-Score of $90(\pm0.5)\%$.

\subsubsection{Driver Demographics}
Next, we evaluate the Fused-CNN model on a personalized scale across all five drivers. In Table~\ref{table:userwise}, we report the action-wise F1-Score of the model. We see consistently good performance across all five drivers, achieving an overall weighted F1-Score of $91(\pm 1)\%$. It is worth noting that specific activities are harder to detect for some drivers due to their height, sitting stances, and sitting positions. Moreover, driving behaviors vary from one driver to another. Despite the driver-specific variations, \ourmethod{} generalizes well and shows significant performance across all drivers.

\begin{table}
\centering
\scriptsize
\caption{Driver-wise weighted F1-Score}
\label{table:userwise}
\begin{tabular}{@{}lccccc@{}}
\toprule
\textbf{Driving Behavior}               & \textbf{Driver-1} & \textbf{Driver-2} & \textbf{Driver-3} & \textbf{Driver-4} & \textbf{Driver-5} \\ \midrule
\textbf{Drinking}            & 0.90            & 0.96            & 0.80            & 0.88            & 0.92            \\
\textbf{Fetching forward}    & 0.94            & 0.86            & 0.91            & 0.94            & 0.87            \\
\textbf{Anomaly in steering} & 0.96            & 0.91            & 0.98            & 0.91            & 0.93            \\
\textbf{Nodding}             & 0.95            & 0.89            & 0.97            & 0.95            & 0.95            \\
\textbf{Yawning}             & 0.96            & 0.86            & 0.81            & 0.87            & 0.88            \\
\textbf{Picking drops}       & 0.96            & 0.85            & 0.87            & 0.82            & 0.84            \\
\textbf{Using phone}         & 0.91            & 0.95            & 0.93            & 0.92            & 0.93            \\
\textbf{Turning back}        & 0.91            & 0.95            & 0.92            & 0.94            & 0.91            \\
\textbf{Talking left}        & 0.90            & 0.86            & 0.89            & 0.92            & 0.93            \\
\textbf{Overall}        & 0.93            & 0.90            & 0.89            & 0.91            & 0.92            \\ \bottomrule
\end{tabular}
\end{table}

    

\subsubsection{Performance of acoustic-FMCW} 
In \figurename~\ref{fig:acoustic_driving}, we show the confusion matrix of classifying the nine dangerous driving behaviors using acoustic-FMCW~\cite{jiang2021driversonar}. For the given classification, it achieves an accuracy of $36\%$, failing miserably as it starts confusing each of the dangerous driving behaviors. For example, \textit{fetching forward} is misclassified as \textit{picking/drops} and \textit{nodding} due to the similar body movement of the driver. Moreover, for dangerous and normal driving classification, the acoustic-FMCW approach shows an F1-Score of $60\%$ and $73\%$, respectively. Thus existing acoustic-based approaches may perform moderately on small-scale behavior classes but remain infeasible for nine different driving behaviors.

\subsubsection{Resource and Energy Consumption}
\begin{figure}
    \centering
    \subfigure[]{
    \includegraphics[width=0.22\textwidth]{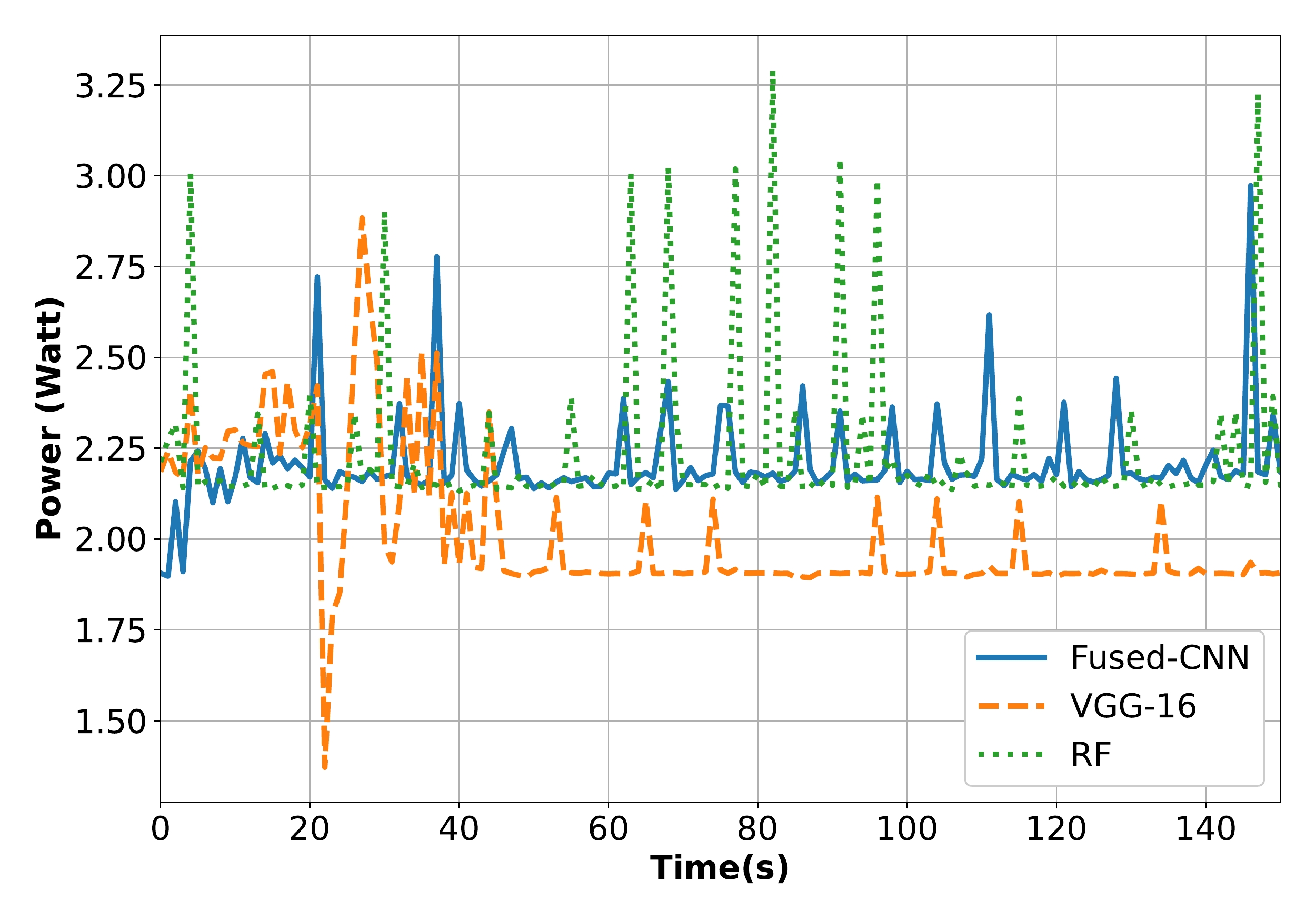}
    \label{fig:power}}
    \subfigure[]{
    \includegraphics[width=0.22\textwidth]{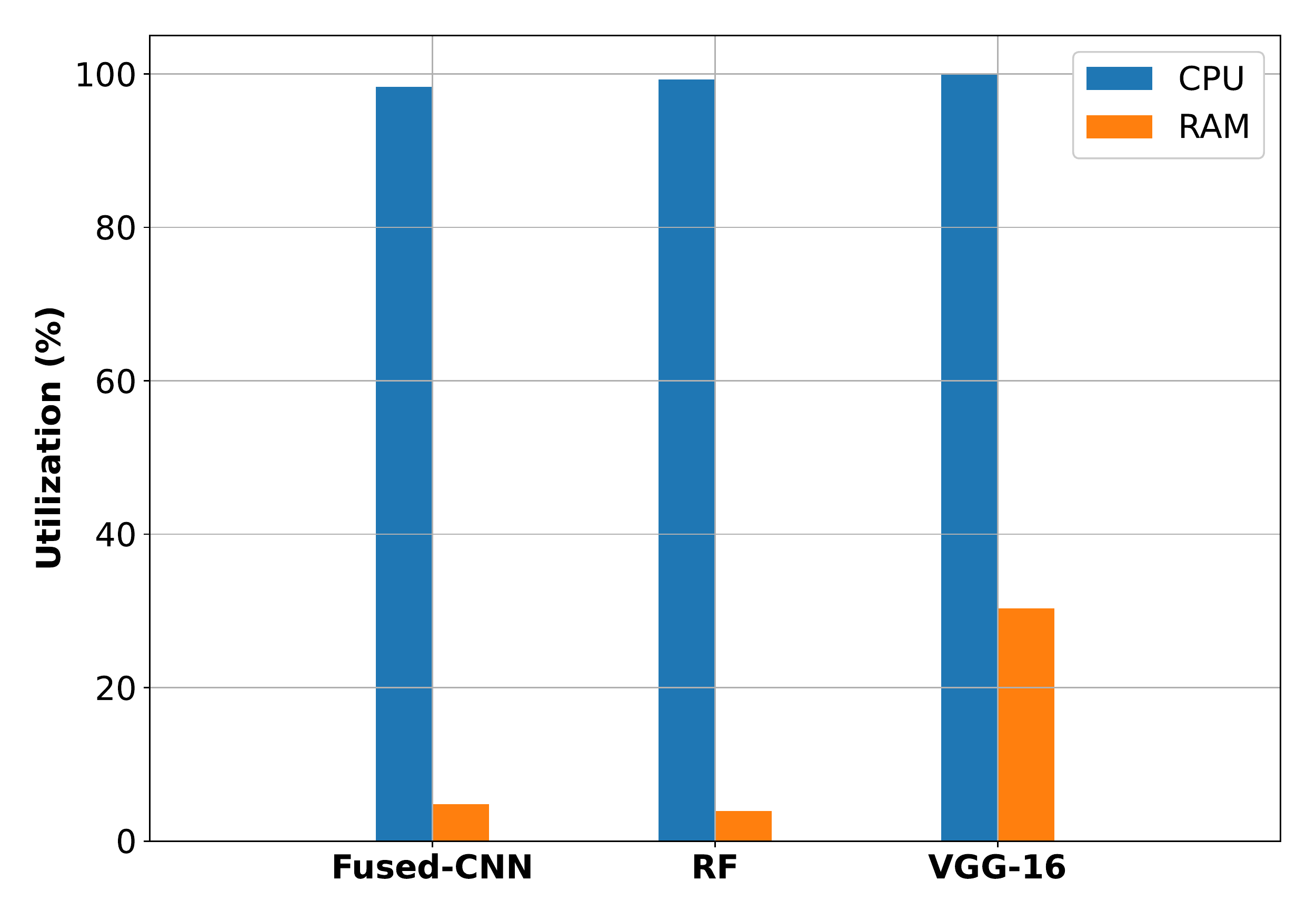}
    \label{fig:res_bar}}
    \caption{(a) Power Consumption, (b) Resource Utilization}
    \label{fig:pow_res}
\end{figure}

Finally, in \figurename~\ref{fig:pow_res}, we report the power consumption footprints as well as the CPU and memory utilization of Fused-CNN of \ourmethod{} with the baselines. As shown in \figurename~\ref{fig:power}, we observe that both Fused-CNN and RF show higher power consumption compared to VGG-16. Due to less computational overhead, RF and Fused-CNN maintain a better utilization of CPU and memory resources, showing lower latency for real-time inference. The peaks in \figurename~\ref{fig:power} represent starting of an inference. However, a higher latency for VGG-16 compared to Fused-CNN and RF-based classifiers indicate that it is unsuitable for live deployment. We further observe the memory usage of all classifiers and find that Fused-CNN consumes an almost similar amount of resources in terms of CPU or memory to RF while having significantly higher accuracy.

\section{Conclusion}
With the increasing demand for on-road safety, dangerous driving and lack of attention while driving continue to be a subject that needs significant attention. For decades, this problem has encouraged researchers to explore solutions that are efficient, pervasive, and, more importantly, \textit{timely}. In this paper, we demonstrate how carefully selecting features from the measurements of a single COTS mmWave FMCW radar could be the most prominent all-around solution to solve this problem. Our advanced solution called \ourmethod is not just compact, pervasive, completely on-device, and privacy-preserving but also demonstrates an accuracy of $\geq95\%$ while detecting a critical set of driver activities that could potentially signify dangerous driving scenarios. We also thoroughly evaluate \ourmethod in several real-world environments and compare its performance with notable baselines. With the encouraging observations of the outcome, we firmly believe \ourmethod could play a crucial role in saving lives and contributing to on-road safety under diverse scenarios.

%
\bibliographystyle{IEEEtran}
\bibliography{refs}

\begin{thebibliography}{10}
\providecommand{\url}[1]{#1}
\csname url@samestyle\endcsname
\providecommand{\newblock}{\relax}
\providecommand{\bibinfo}[2]{#2}
\providecommand{\BIBentrySTDinterwordspacing}{\spaceskip=0pt\relax}
\providecommand{\BIBentryALTinterwordstretchfactor}{4}
\providecommand{\BIBentryALTinterwordspacing}{\spaceskip=\fontdimen2\font plus
\BIBentryALTinterwordstretchfactor\fontdimen3\font minus
  \fontdimen4\font\relax}
\providecommand{\BIBforeignlanguage}[2]{{%
\expandafter\ifx\csname l@#1\endcsname\relax
\typeout{** WARNING: IEEEtran.bst: No hyphenation pattern has been}%
\typeout{** loaded for the language `#1'. Using the pattern for}%
\typeout{** the default language instead.}%
\else
\language=\csname l@#1\endcsname
\fi
#2}}
\providecommand{\BIBdecl}{\relax}
\BIBdecl

\bibitem{world2020road}
\BIBentryALTinterwordspacing
I.~T. Forum, \emph{Road Safety Annual Report 2020}.\hskip 1em plus 0.5em minus
  0.4em\relax OECD, 2020, (Last accessed: October 10, 2022). [Online].
  Available:
  \url{https://www.oecd-ilibrary.org/content/publication/f3e48023-en}
\BIBentrySTDinterwordspacing

\bibitem{shaout2011advanced}
A.~Shaout, D.~Colella, and S.~Awad, ``Advanced driver assistance systems-past,
  present and future,'' in \emph{2011 Seventh International Computer
  Engineering Conference}.\hskip 1em plus 0.5em minus 0.4em\relax IEEE, 2011,
  pp. 72--82.

\bibitem{yin2017automatic}
J.-L. Yin, B.-H. Chen, K.-H.~R. Lai, and Y.~Li, ``Automatic dangerous driving
  intensity analysis for advanced driver assistance systems from multimodal
  driving signals,'' \emph{IEEE Sensors Journal}, vol.~18, no.~12, pp.
  4785--4794, 2017.

\bibitem{jiang2021smart}
L.~Jiang, W.~Xie, D.~Zhang, and T.~Gu, ``Smart diagnosis: Deep learning boosted
  driver inattention detection and abnormal driving prediction,'' \emph{IEEE
  Internet of Things Journal}, vol.~9, no.~6, pp. 4076--4089, 2021.

\bibitem{yang2020driver}
H.~Yang, L.~Liu, W.~Min, X.~Yang, and X.~Xiong, ``Driver yawning detection
  based on subtle facial action recognition,'' \emph{IEEE Transactions on
  Multimedia}, vol.~23, pp. 572--583, 2020.

\bibitem{cao2022towards}
Y.~Cao, F.~Li, X.~Liu, S.~Yang, and Y.~Wang, ``Towards reliable driver
  drowsiness detection leveraging wearables,'' \emph{ACM Transactions on Sensor
  Networks}, 2022.

\bibitem{wang2021survey}
J.~Wang, W.~Chai, A.~Venkatachalapathy, K.~L. Tan, A.~Haghighat,
  S.~Velipasalar, Y.~Adu-Gyamfi, and A.~Sharma, ``A survey on driver behavior
  analysis from in-vehicle cameras,'' \emph{IEEE Transactions on Intelligent
  Transportation Systems}, 2021.

\bibitem{kundinger2020feasibility}
T.~Kundinger, P.~K. Yalavarthi, A.~Riener, P.~Wintersberger, and
  C.~Schartm{\"u}ller, ``Feasibility of smart wearables for driver drowsiness
  detection and its potential among different age groups,'' \emph{International
  Journal of Pervasive Computing and Communications}, 2020.

\bibitem{hur2016proposal}
S.~Hur, S.~Baek, B.~Kim, Y.~Chang, A.~F. Molisch, T.~S. Rappaport, K.~Haneda,
  and J.~Park, ``Proposal on millimeter-wave channel modeling for 5g cellular
  system,'' \emph{IEEE journal of selected topics in signal processing},
  vol.~10, no.~3, pp. 454--469, 2016.

\bibitem{li2020capturing}
G.~Li, Z.~Zhang, H.~Yang, J.~Pan, D.~Chen, and J.~Zhang, ``Capturing human pose
  using mmwave radar,'' in \emph{2020 IEEE International Conference on
  Pervasive Computing and Communications Workshops (PerCom Workshops)}.\hskip
  1em plus 0.5em minus 0.4em\relax IEEE, 2020, pp. 1--6.

\bibitem{wang2021m}
Y.~Wang, H.~Liu, K.~Cui, A.~Zhou, W.~Li, and H.~Ma, ``m-activity: Accurate and
  real-time human activity recognition via millimeter wave radar,'' in
  \emph{ICASSP 2021-2021 IEEE International Conference on Acoustics, Speech and
  Signal Processing}.\hskip 1em plus 0.5em minus 0.4em\relax IEEE, 2021, pp.
  8298--8302.

\bibitem{palipana2021pantomime}
S.~Palipana, D.~Salami, L.~A. Leiva, and S.~Sigg, ``Pantomime: Mid-air gesture
  recognition with sparse millimeter-wave radar point clouds,''
  \emph{Proceedings of the ACM on Interactive, Mobile, Wearable and Ubiquitous
  Technologies}, vol.~5, no.~1, pp. 1--27, 2021.

\bibitem{yang2016monitoring}
Z.~Yang, P.~H. Pathak, Y.~Zeng, X.~Liran, and P.~Mohapatra, ``Monitoring vital
  signs using millimeter wave,'' in \emph{Proceedings of the 17th ACM
  international symposium on mobile ad hoc networking and computing}, 2016, pp.
  211--220.

\bibitem{liu2021wavoice}
T.~Liu, M.~Gao, F.~Lin, C.~Wang, Z.~Ba, J.~Han, W.~Xu, and K.~Ren, ``Wavoice: A
  noise-resistant multi-modal speech recognition system fusing mmwave and audio
  signals,'' in \emph{Proceedings of the 19th ACM Conference on Embedded
  Networked Sensor Systems}, 2021, pp. 97--110.

\bibitem{sen2023mmassist}
A.~Sen, A.~Das, P.~Karmakar, and S.~Chakraborty, ``mmassist: Passive monitoring
  of driver's attentiveness using mmwave sensors,'' in \emph{2023 15th
  International Conference on COMmunication Systems \& NETworkS
  (COMSNETS)}.\hskip 1em plus 0.5em minus 0.4em\relax IEEE, 2023, pp. 545--553.

\bibitem{liu2022remote}
S.~Liu, L.~Zhao, X.~Yang, Y.~Du, M.~Li, X.~Zhu, and Z.~Dai, ``Remote drowsiness
  detection based on the mmwave fmcw radar,'' \emph{IEEE Sensors Journal},
  vol.~22, no.~15, pp. 15\,222--15\,234, 2022.

\bibitem{zhao2020heart}
P.~Zhao, C.~X. Lu, B.~Wang, C.~Chen, L.~Xie, M.~Wang, N.~Trigoni, and
  A.~Markham, ``Heart rate sensing with a robot mounted mmwave radar,'' in
  \emph{2020 IEEE International Conference on Robotics and Automation
  (ICRA)}.\hskip 1em plus 0.5em minus 0.4em\relax IEEE, 2020, pp. 2812--2818.

\bibitem{jiang2021driversonar}
H.~Jiang, J.~Hu, D.~Liu, J.~Xiong, and M.~Cai, ``Driversonar: Fine-grained
  dangerous driving detection using active sonar,'' \emph{Proceedings of the
  ACM on Interactive, Mobile, Wearable and Ubiquitous Technologies}, vol.~5,
  no.~3, pp. 1--22, 2021.

\bibitem{ersal2010model}
T.~Ersal, H.~J. Fuller, O.~Tsimhoni, J.~L. Stein, and H.~K. Fathy,
  ``Model-based analysis and classification of driver distraction under
  secondary tasks,'' \emph{IEEE transactions on intelligent transportation
  systems}, vol.~11, no.~3, pp. 692--701, 2010.

\bibitem{dai2010mobile}
J.~Dai, J.~Teng, X.~Bai, Z.~Shen, and D.~Xuan, ``Mobile phone based drunk
  driving detection,'' in \emph{2010 4th International Conference on Pervasive
  Computing Technologies for Healthcare}.\hskip 1em plus 0.5em minus
  0.4em\relax IEEE, 2010, pp. 1--8.

\bibitem{cyganek2014hybrid}
B.~Cyganek and S.~Gruszczy{\'n}ski, ``Hybrid computer vision system for
  drivers' eye recognition and fatigue monitoring,'' \emph{Neurocomputing},
  vol. 126, pp. 78--94, 2014.

\bibitem{borghi2017poseidon}
G.~Borghi, M.~Venturelli, R.~Vezzani, and R.~Cucchiara, ``Poseidon:
  Face-from-depth for driver pose estimation,'' in \emph{Proceedings of the
  IEEE conference on computer vision and pattern recognition}, 2017, pp.
  4661--4670.

\bibitem{dua2020dgaze}
I.~Dua, T.~A. John, R.~Gupta, and C.~Jawahar, ``Dgaze: Driver gaze mapping on
  road,'' in \emph{2020 IEEE/RSJ International Conference on Intelligent Robots
  and Systems}.\hskip 1em plus 0.5em minus 0.4em\relax IEEE, 2020, pp.
  5946--5953.

\bibitem{kajiwara2021driver}
S.~Kajiwara, ``Driver-condition detection using a thermal imaging camera and
  neural networks,'' \emph{International journal of automotive technology},
  vol.~22, no.~6, pp. 1505--1515, 2021.

\bibitem{yan2011robust}
C.~Yan, Y.~Wang, and Z.~Zhang, ``Robust real-time multi-user pupil detection
  and tracking under various illumination and large-scale head motion,''
  \emph{Computer Vision and Image Understanding}, vol. 115, no.~8, pp.
  1223--1238, 2011.

\bibitem{hoang2016multiple}
T.~Hoang Ngan~Le, Y.~Zheng, C.~Zhu, K.~Luu, and M.~Savvides, ``Multiple scale
  faster-rcnn approach to driver's cell-phone usage and hands on steering wheel
  detection,'' in \emph{Proceedings of the IEEE conference on computer vision
  and pattern recognition workshops}, 2016, pp. 46--53.

\bibitem{zhang2016privacy}
L.~Zhang, K.~Liu, X.-Y. Li, C.~Liu, X.~Ding, and Y.~Liu, ``Privacy-friendly
  photo capturing and sharing system,'' in \emph{Proceedings of the 2016 ACM
  International Joint Conference on Pervasive and Ubiquitous Computing}, 2016,
  pp. 524--534.

\bibitem{cisotto2018joint}
G.~Cisotto, A.~V. Guglielmi, L.~Badia, and A.~Zanella, ``Joint compression of
  eeg and emg signals for wireless biometrics,'' in \emph{2018 IEEE Global
  Communications Conference}.\hskip 1em plus 0.5em minus 0.4em\relax IEEE,
  2018, pp. 1--6.

\bibitem{xu2017er}
X.~Xu, H.~Gao, J.~Yu, Y.~Chen, Y.~Zhu, G.~Xue, and M.~Li, ``Er: Early
  recognition of inattentive driving leveraging audio devices on smartphones,''
  in \emph{IEEE INFOCOM 2017-IEEE Conference on Computer Communications}.\hskip
  1em plus 0.5em minus 0.4em\relax IEEE, 2017, pp. 1--9.

\bibitem{xie2019d}
Y.~Xie, F.~Li, Y.~Wu, S.~Yang, and Y.~Wang, ``D 3-guard: Acoustic-based drowsy
  driving detection using smartphones,'' in \emph{IEEE INFOCOM 2019-IEEE
  Conference on Computer Communications}.\hskip 1em plus 0.5em minus
  0.4em\relax IEEE, 2019, pp. 1225--1233.

\bibitem{bai2020acoustic}
Y.~Bai, L.~Lu, J.~Cheng, J.~Liu, Y.~Chen, and J.~Yu, ``Acoustic-based sensing
  and applications: A survey,'' \emph{Computer Networks}, vol. 181, p. 107447,
  2020.

\bibitem{hromadova2022frequency}
V.~Hromadov{\'a}, P.~Kas{\'a}k, R.~Jarina, and P.~Br{\'\i}da, ``Frequency
  response of smartphones at the upper limit of the audible range,'' in
  \emph{2022 ELEKTRO}.\hskip 1em plus 0.5em minus 0.4em\relax IEEE, 2022, pp.
  1--5.

\bibitem{3478084}
\BIBentryALTinterwordspacing
H.~Jiang, J.~Hu, D.~Liu, J.~Xiong, and M.~Cai, ``Driversonar: Fine-grained
  dangerous driving detection using active sonar,'' \emph{Proc. ACM Interact.
  Mob. Wearable Ubiquitous Technol.}, vol.~5, no.~3, sep 2021. [Online].
  Available: \url{https://doi.org/10.1145/3478084}
\BIBentrySTDinterwordspacing

\bibitem{mandal2021exploiting}
R.~Mandal, P.~Karmakar, S.~Chatterjee, D.~D. Spandan, S.~Pradhan, S.~Saha,
  S.~Chakraborty, and S.~Nandi, ``Exploiting multi-modal contextual sensing for
  city-bus's stay location characterization: Towards sub-60 seconds accurate
  arrival time prediction,'' \emph{ACM Transactions on Internet Things}, 2022.

\bibitem{wang2016interacting}
S.~Wang, J.~Song, J.~Lien, I.~Poupyrev, and O.~Hilliges, ``Interacting with
  soli: Exploring fine-grained dynamic gesture recognition in the
  radio-frequency spectrum,'' in \emph{Proceedings of the 29th Annual Symposium
  on User Interface Software and Technology}, 2016, pp. 851--860.

\bibitem{liu2021m}
H.~Liu, A.~Zhou, Z.~Dong, Y.~Sun, J.~Zhang, L.~Liu, H.~Ma, J.~Liu, and N.~Yang,
  ``M-gesture: Person-independent real-time in-air gesture recognition using
  commodity millimeter wave radar,'' \emph{IEEE Internet of Things Journal},
  vol.~9, no.~5, pp. 3397--3415, 2021.

\bibitem{tiwari2021mmwave}
G.~Tiwari and S.~Gupta, ``An mmwave radar based real-time contactless fitness
  tracker using deep cnns,'' \emph{IEEE Sensors Journal}, vol.~21, no.~15, pp.
  17\,262--17\,270, 2021.

\bibitem{simonyan2014very}
K.~Simonyan and A.~Zisserman, ``Very deep convolutional networks for
  large-scale image recognition,'' \emph{arXiv preprint arXiv:1409.1556}, 2014.

\bibitem{koehrsen2018hyperparameter}
W.~Koehrsen, ``Hyperparameter tuning the random forest in python,''
  \emph{Towards Data Science}, 2018.

\bibitem{deng2009imagenet}
J.~Deng, W.~Dong, R.~Socher, L.-J. Li, K.~Li, and L.~Fei-Fei, ``Imagenet: A
  large-scale hierarchical image database,'' in \emph{2009 IEEE conference on
  computer vision and pattern recognition}.\hskip 1em plus 0.5em minus
  0.4em\relax Ieee, 2009, pp. 248--255.

\bibitem{xie2020real}
Y.~Xie, F.~Li, Y.~Wu, S.~Yang, and Y.~Wang, ``Real-time detection for drowsy
  driving via acoustic sensing on smartphones,'' \emph{IEEE Transactions on
  Mobile Computing}, vol.~20, no.~8, pp. 2671--2685, 2020.

\end{thebibliography}
\end{document}